\documentclass{jfm}
\usepackage{graphicx}
\usepackage{newtxtext}
\usepackage{newtxmath}
\usepackage{natbib}
\usepackage{hyperref}
\usepackage[dvipsnames]{xcolor}
\usepackage{cancel}
\usepackage{pifont}
\usepackage{relsize}
\usepackage{tikz}
\usetikzlibrary{tikzmark}
\tikzset{mycircled/.style={circle,draw,inner sep=0.1em,line width=0.1em}}
\hypersetup{
    colorlinks = true,
    urlcolor   = blue,
    citecolor  = blue,
}
\usepackage{xcolor}

\newcommand{\RomanNumeralCaps}[1]
\linenumbers

\shorttitle{Expansion of a hole in a viscoelastic liquid sheet}
\shortauthor{T. Ruangkriengsin, R. Brandão, and H.A. Stone}

\title{\Large Expansion of a hole in a viscoelastic liquid sheet}
 \author{Tachin Ruangkriengsin\aff{1},
Rodolfo Brandão\aff{2}, \and 
Howard A. Stone\aff{3}
 \corresp{\email\href{mailto:hastone@princeton.edu}{hastone@princeton.edu}}
}
  \affiliation{\aff{1}Program in Applied and Computational Mathematics, Princeton University, Princeton, NJ 08544, USA
  \aff{2}Department of Mathematics, University of British Columbia, Vancouver, BC V6T 1Z2
  \aff{3}Department of Mechanical and Aerospace Engineering, Princeton University, Princeton, NJ 08544, USA}
   
\begin{document}
\maketitle
\begin{abstract}
Experiments on highly viscous polymeric films show that punctured holes expand exponentially in time, without sustained accumulation of liquid near the rim. This response departs from the Taylor--Culick description, in which displaced liquid accumulates in a growing rim that moves at constant speed. Although these differences were initially attributed to viscoelasticity, they were later rationalized using a purely viscous theory, leaving the role of viscoelastic stresses unresolved.
We analyze the expansion of an axisymmetric hole in a freely suspended viscoelastic liquid sheet described by the Oldroyd-B model. Exploiting the separation of length scales between hole radius and film thickness, we derive extensional thin-film equations on the scale of the hole and an effective boundary condition from an asymptotic force balance in the tip region. Analytical solutions 
are obtained for weak viscoelasticity, $Wi\ll 1$, and the ultra-dilute limit, $\mu_p\ll\mu_s$, where $Wi$ is the Weissenberg number, while $\mu_s$ and $\mu_p$ are solvent and polymeric viscosities, respectively. 
For weak viscoelasticity, the dimensionless hole radius grows approximately as $e^{(0.5+\alpha Wi \beta_p)T}$, where $\alpha=(12 - 6\log 2-\pi)/21\approx 0.224$ and $\beta_p=\mu_p/(\mu_s+\mu_p)$. 
In the ultra-dilute limit, the radius grows approximately as $e^{(0.5+\alpha^{*}\beta_p )T}$, where $\alpha^{*}(Wi)>0$ is evaluated numerically.
In both regimes, viscoelastic stresses increase the exponential growth rate relative to the Newtonian limit and induce film-thickness variations, with thickening near the retracting edge.
This acceleration arises from azimuthal stretching and radial compression of the polymers, which redistribute stresses in the film and modify the stress balance at the tip, leading to a stronger outward radial extensional flow.

\end{abstract}

\section{Introduction}

The expansion of a hole in a punctured liquid film is a capillary-driven phenomenon central to breakup processes such as bubble bursting and aerosol production at liquid--gas interfaces~\citep{Bremond:05,Lhuissier:12, veron2015ocean, deike2022mass} and sheet fragmentation in sprays and atomization~\citep{villermaux:02,bremond:07,eggers:08,Villermaux:20, eshima2025precursors}. Although the simplest cases involve Newtonian films surrounded by air, many practical films contain polymers, surfactants, or other microstructural components. In such systems, the retraction dynamics can be altered by viscous, elastic, or interfacial stresses~\citep{Debregeas:95,Debregeas:98,villone:17,tammaro:18,constante2022role}.

The classical theory for the inertial retraction of a hole in an infinite, flat Newtonian sheet was developed independently by~\citet{Taylor:59} and~\citet{Culick:60}, building on earlier ideas of \citet{Dupre:67} and \citet{Rayleigh:91}. In this regime, fluid accumulates in a circular rim that propagates outward and approaches a constant speed at sufficiently long times. This predicted speed, now commonly known as the Taylor--Culick velocity, was later shown to agree well with experiments using low-viscosity fluids~\citep{Ranz:59, McEntee:69}. Beyond this ideal setting, the framework was extended to non-uniform liquid films and threads~\citep{Keller:83}.

Later experiments by~\citet{Debregeas:95, Debregeas:98}, however, revealed different dynamics for highly viscous polymeric films. Rather than accumulating fluid near the retracting edge, punctured films remained approximately uniform in thickness while the hole radius grew exponentially in time. Similar exponential growth was also reported in molten polystyrene films~\citep{dalnoki:99}. These observations, which differed from those predicted by Taylor--Culick theory, were originally attributed to the viscoelastic response of the polymeric liquid. In particular, \citet{Debregeas:95} proposed that surface-tension forces acting at the film edge are elastically propagated into the film at a velocity set by the shear modulus of the material, thereby inhibiting thickening near the edge and allowing the film to remain flat during retraction.

This interpretation was subsequently revisited by~\citet{Brenner:99}, who demonstrated numerically, for planar configurations, that the absence of a localized increase in thickness near the retracting edge can occur even in Newtonian films provided
viscous effects dominate inertial effects. An analytical thin-film framework for this Newtonian viscous regime was later developed by~\citet{Savva:09}, who analyzed both the planar sheet and the axisymmetric hole problems, showing that the hole radius grows exponentially at early times in the latter geometry. Subsequent studies refined the planar viscous-sheet problem by resolving its long-time asymptotic structure~\citep{Gordillo:11} and incorporating finite-length effects~\citep{Deka:20}.

Although~\citet{Brenner:99} and~\citet{Savva:09} provided important theoretical descriptions of viscous-film retraction, their analyses assumed the thin-film approximation to remain valid throughout the liquid domain, including near the highly curved retracting edge---an assumption that~\citet{Brenner:99} noted was ``not strictly asymptotically correct.'' Furthermore, in deriving the exponential expansion of an axisymmetric hole,~\citet{Savva:09} considered an initial profile consisting of a horizontal strip connected to a semicircular arc. While this profile provides a convenient initial configuration, it introduces a curvature discontinuity at the junction between the uniform film and the semicircle, which must then be accounted for explicitly through a jump condition.

An alternative description of the problem, avoiding both the use of thin-film theory at the curved tip and the introduction of curvature discontinuities, was proposed by~\citet{Munro:Thesis}. There, the liquid domain is decomposed into a thin-film region and a tip region, whose asymptotic matching yields an effective boundary condition for the thin-film problem. Recently,~\citet{ahsan-rodolfo:26} adopted this decomposition to study the retraction of a planar sheet, showing that it is asymptotically valid when the Ohnesorge number $Oh$, which measures the relative importance of viscous to inertial effects, is large, $Oh\gg 1$, and the elapsed time after rupture remains sufficiently small compared with a long inertial timescale.

In the present work, we employ an asymptotic approach similar to that of~\citet{ahsan-rodolfo:26} to study the expansion of an axisymmetric hole in a thin viscoelastic film modeled as an Oldroyd-B fluid, focusing on the large-$Oh$ regime. In the planar geometry considered by~\citet{ahsan-rodolfo:26}, inertia must be retained to regularize the far-field velocity. In the present axisymmetric geometry, as we shall see, radial spreading allows the velocity to decay in the far field, so that inertia can be neglected everywhere.

A particular objective of our analysis is to examine how viscoelasticity modifies the retraction dynamics.
Although previous purely viscous theories explain many qualitative features of experiments by~\citet{Debregeas:95}, including the suppressed rim formation and exponential growth of the hole, some discrepancies have been reported. In particular,~\citet{Savva:09} predict that the hole radius expands with a growth rate roughly $30\%$ below the experimental values reported by~\citet{Debregeas:95} using polymeric films. While this discrepancy could partly reflect details of the puncture or the initial film profile, it also suggests that polymeric stresses, including those associated with normal-stress differences, may enhance the retraction rate.

Consistent with this possibility, recent numerical and experimental studies have shown that viscoelasticity modifies the capillary retraction of liquid films and filaments. Direct numerical simulations of viscoelastic films found that film retraction depends on the constitutive model and deformation history of the polymeric liquid~\citep{villone:17,villone:19}, while experiments on the relaxation of viscoelastic filaments showed that stretched polymers can provide an additional elastic tension that increases the retraction speed~\citep{sen:21}. Related experiments on bursting viscoelastic bubbles also revealed that stored elastic stresses can affect rupture dynamics and produce distinct morphologies, such as the flowering instability of polymeric bubbles~\citep{tammaro:18,tammaro:21}. These studies provide evidence that viscoelastic stresses can modify capillary retraction in several geometries, but a systematic mathematical framework isolating their effect on the expansion of an axisymmetric hole in a thin film remains lacking.

The remaining sections of this paper are organized as follows. In $\mathsection$\ref{formulation}, we formulate the problem and identify the scalings in the thin-film and tip regions. In $\mathsection$\ref{hole_setup}, we derive the extensional thin-film equations for the Oldroyd-B model, which govern the dynamics on the scale of the hole. In $\mathsection$\ref{tip_setup}, we obtain the effective boundary condition at the retracting edge by matching to the force balance in the tip region. In $\mathsection$\ref{derivation_section}, we present the reduced model and introduce a coordinate transformation to a frame moving with the hole edge. We then consider two asymptotic limits. In $\mathsection$\ref{small-wi-section}, we analyze the weakly viscoelastic regime, $Wi\ll1$, and obtain the solution analytically to first order in $Wi$. In $\mathsection$\ref{ultra-dilute-section}, we examine the ultra-dilute regime, $\beta_p\ll1$, and derive the corresponding solution to first order in $\beta_p$. Finally, in $\mathsection$\ref{conclusion}, we summarize the main results and discuss their implications.

\section{Problem formulation}\label{formulation}

\subsection{Problem statement}

\begin{figure}
    \centering
    \includegraphics[width= 0.75\columnwidth]{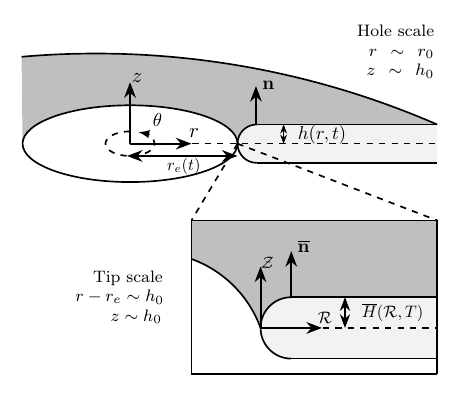}
    \caption{Schematic illustration of an expanding hole in a viscoelastic liquid sheet.}
    \label{schematic}
\end{figure}

We consider the opening of an axisymmetric hole in a thin viscoelastic liquid film, as illustrated in figure~\ref{schematic}. We employ cylindrical coordinates $(r,\theta,z)$, with corresponding unit vectors $(\textbf{e}_r,\textbf{e}_{\theta},\textbf{e}_z)$, and place the origin at the center of the hole. We denote the radius of the hole by $r_e(t)$, for time $t \geq 0$. The liquid, with velocity field $\textbf{u}(r,z,t)=u_r(r,z,t)\textbf{e}_r+u_z(r,z,t)\textbf{e}_z$, is suspended between two free surfaces $z=\pm h(r,t)$, defined for $r\in[r_e(t),\infty)$. Here, we assume symmetry about the midplane $z=0$.

We take the initial hole radius to be $r_e(0)=r_0$ and assume a uniform initial film thickness $h(r,0)=h_0$, except within a narrow annular region near the edge of the hole. We focus on the regime in which the film thickness is small compared with the hole size,
\begin{equation}
\epsilon = \frac{h_0}{r_0} \ll 1.
\label{epsilon}
\end{equation}
In this regime, the initial film thickness can be expressed as
\begin{equation}
h(r,0) = h_0 \quad \text{for} \quad \frac{r}{r_0}-1 \gg \epsilon.
\label{initial_thickness}
\end{equation}
Together, \eqref{epsilon} and \eqref{initial_thickness} motivate a decomposition of the problem into two regions: a hole-scale region, in which the radial length scale is set by the initial hole radius and the vertical length scale by the film thickness, and a tip-scale region, in which both radial and vertical length scales are of order the film thickness (see figure~\ref{schematic}).

We restrict attention to the hole-scale region, where the initial condition \eqref{initial_thickness} applies. In this region, a thin-film (long-wavelength) approximation is applicable, leading to a simplified description of the hole-scale dynamics. Although no initial condition is prescribed on the tip scale, an effective boundary condition for the hole-scale problem can be obtained through a global force balance in the tip-scale region \citep{ahsan-rodolfo:26}. Details of the hole- and tip-scale analyses are provided in $\mathsection$$\mathsection$~\ref{hole_setup} and~\ref{tip_setup}, respectively.

To model the viscoelastic contributions in the film, we adopt the Oldroyd-B constitutive description~\citep{oldroyd:50, bird:87}, for which the total stress tensor is given by
\begin{equation}\label{total_stress}
    \boldsymbol{\sigma} = -p \mathbf{I} + 2\mu_s \mathbf{E} + \boldsymbol{\sigma}_p.
\end{equation}
Here, the first term represents the pressure, while the second term corresponds to the viscous contribution from the solvent of viscosity $\mu_s$, with $\mathbf{E} = \tfrac{1}{2}\left(\nabla \mathbf{u} + (\nabla \mathbf{u})^{T}\right)$ denoting the rate-of-strain tensor. The third term represents the polymeric contribution to the stress, which evolves according to
\begin{equation}\label{polymer_stress}
    \boldsymbol{\sigma}_p + \lambda\left(\frac{\partial \boldsymbol{\sigma}_p}{\partial t} + \textbf{u} \cdot \bnabla \boldsymbol{\sigma}_p - \boldsymbol{\sigma}_p \cdot \bnabla \textbf{u} - (\bnabla \textbf{u})^{\rm T}\cdot \boldsymbol{\sigma}_p\right) = 2\mu_p\textbf{E},
\end{equation}
where $\lambda$ and $\mu_p$ represent the relaxation time and viscosity of the polymeric solution, respectively. Owing to axisymmetry and the absence of azimuthal flow, the total stress tensor has four independent non-zero components: the three normal stresses $\sigma_{rr}$, $\sigma_{\theta\theta}$ and $\sigma_{zz}$, and the shear stress $\sigma_{rz} = \sigma_{zr}$. We assume that the polymers are initially relaxed, so that
\begin{equation}\label{polymer_stress_init}
    \boldsymbol{\sigma}_p = \boldsymbol{0}\quad \text{at} \quad t=0.
\end{equation}

We consider a highly viscous film, so that inertial effects can be neglected in the present axisymmetric geometry. We clarify the corresponding scaling constraints in $\mathsection\mathsection$~\ref{local_scaling} and \ref{global_scaling}. The continuity equation takes the form
\begin{equation}\label{continuity}
\frac{1}{r}\frac{\partial}{\partial r}\left(ru_r\right) + \frac{\partial u_z}{\partial z} = 0.
\end{equation}
Neglecting inertia, the momentum equation reduces to
\begin{equation}\label{momentum}
\bnabla \cdot \boldsymbol{\sigma} = \mathbf{0}.
\end{equation}

Equations~\eqref{total_stress}--\eqref{momentum} are supplemented by boundary conditions at the free surfaces and in the far field. The dynamic boundary condition enforces a balance between the fluid stress and capillary stress,
\begin{equation}\label{dynamic_bc}
    \mathbf{n} \cdot \boldsymbol{\sigma}= \gamma \kappa \mathbf{n}
    \quad \text{at} \quad z = \pm h(r,t),
\end{equation}
where $\mathbf{n}$ is the unit outward normal to the fluid, $\gamma$ is the surface tension, and $\kappa$ is the total curvature of the interface. The kinematic boundary condition requires the free surfaces to move with the fluid, $\frac{D}{Dt}\bigl(z \mp h(r,t)\bigr)=0$, where $\frac{D}{Dt}$ is the material derivative, which can be expressed as
\begin{equation}\label{kinematic_bc}
u_z = \pm\left(\frac{\partial h}{\partial t}
+ u_r \frac{\partial h}{\partial r}\right)
\quad \text{at} \quad z = \pm h(r,t).
\end{equation}
Equation~\eqref{kinematic_bc} is valid away from the tip region near $r = r_e(t)$. At the tip, the kinematic condition requires the hole edge to move with the local fluid velocity, $\frac{D}{Dt}\bigl(r - r_e(t)\bigr) = 0$,
or
\begin{equation}\label{kinematic_tip}
\frac{dr_e}{dt} = u_r
\quad \text{at} \quad r = r_e(t), \quad z = 0.
\end{equation}
Lastly, we assume that the fluid velocity decays to zero in the far field, 
\begin{equation}\label{far_field}
    \mathbf{u} \to \mathbf{0} \quad \text{as }\quad r \to \infty.
\end{equation}

The boundary conditions above assume an inviscid external medium and, without loss of generality, we take the ambient pressure to be zero, $p_{\mathrm{amb}}=0$. The same assumptions have been made in analyses of Newtonian film retraction \citep{Brenner:99,Savva:09,Deka:20,ahsan-rodolfo:26}. Recently, \citet{Sanjay:22} studied the retraction of a sheet in an external medium of comparable viscosity.

We preface the analysis of the tip and hole regions, in the limit $\epsilon\ll1$, by estimating the characteristic magnitudes of the flow and stress fields in each region.

\subsection{Scalings in the tip-scale region}\label{local_scaling}

The expansion of the hole is driven by the large capillary stresses arising from the strong curvature of the interface near the tip of the film. In this region, the characteristic length scale is set by the film thickness $h_0$, so that the dominant curvature scales as $\kappa \sim h_0^{-1}$. Balancing viscous and capillary stresses in \eqref{dynamic_bc} then yields a characteristic local velocity scale in the tip region,
\begin{equation}\label{tip_velocity}
u_{\text{tip}} \sim \frac{\gamma}{\mu},
\end{equation}
where $\mu = \mu_s + \mu_p$ is the total viscosity. As we shall see, this is \emph{not} the same as the scale of the hole-expansion speed $d r_e/dt$. This scaling also assumes that the polymeric stress is at most comparable to the viscous and capillary stresses; equivalently, the Weissenberg number defined by
\begin{equation}\label{weissenberg}
Wi = \frac{\lambda u_{\text{tip}}}{h_0} = \frac{\lambda \gamma}{\mu h_0}
\end{equation}
is at most $O(1)$. The pressure and stresses in the tip-scale region then scale as $\gamma/h_0$.

To determine when inertia may be neglected locally near the tip, we consider the tip-scale Reynolds number
\begin{equation}
    Re_{\text{tip}} = \frac{\rho u_{\text{tip}} h_0 }{\mu}
    = \frac{\rho \gamma h_0}{\mu^2}
    = \frac{1}{Oh^2},
\end{equation}
where the Ohnesorge number is defined as
\begin{equation}
    Oh = \frac{\mu}{\sqrt{\rho \gamma h_0}}.
\end{equation}
Therefore, inertia in the tip region may be neglected provided that $Oh \gg 1$.

\subsection{Scalings in the hole-scale region}\label{global_scaling}

On the hole scale, the film thickness remains of order $h_0$, whereas the characteristic radial length is now $r_0$. As is typical for freely suspended thin films, the shear-free condition at the free surface---\emph{i.e.}, the tangential component of \eqref{dynamic_bc}---implies a plug-like leading-order flow with the radial velocity varying primarily in the radial direction \citep{Erneux:93, Howell:96}. Denoting the characteristic velocity in the hole-scale region by $U$, the characteristic extensional stress is therefore $\mu U/r_0$. Balancing this with the $O(\gamma/h_0)$ capillary stress generated at the tip gives
\begin{equation}
    U = \frac{\gamma r_0}{\mu h_0}.
\end{equation}
Using the kinematic condition \eqref{kinematic_tip}, a characteristic viscocapillary timescale is $r_0/U = \mu h_0/\gamma$.

The continuity equation then implies that the characteristic vertical velocity in the hole-scale region is $\epsilon U$. Furthermore, the momentum balance \eqref{momentum} suggests that the shear stress $\sigma_{rz}$ is smaller than the normal stresses ($\sigma_{zz}, \sigma_{\theta\theta}, \sigma_{rr}$) by a factor of $\epsilon$, and is therefore of order $\epsilon \mu U/r_0$.

The Reynolds number based on the hole-scale velocity and length is
\begin{equation}
    Re_{\text{film}} = \frac{\rho U r_0}{\mu}
    = \frac{\rho \gamma r_0^2}{\mu^2 h_0}
    = \frac{1}{\epsilon^2 Oh^2}.
\end{equation}
Thus, neglecting inertia on the hole scale requires $Re_{\text{film}}\ll 1$, or equivalently $Oh\gg \epsilon^{-1}$. This is a stronger requirement than that obtained in the tip region; under this condition, inertia is negligible everywhere in the film.

We also note that, since the curvature in this region scales as
$\kappa \sim \nabla^2 h \sim h_0/r_0^2$, the Young--Laplace pressure on the
right-hand side of \eqref{dynamic_bc} scales as $\gamma h_0/r_0^2$. This is
smaller than the extensional viscous stress by a factor of $O(\epsilon^2)$.
Accordingly, while the leading-order flow in the film is driven by capillary
stress originating from the highly curved tip, local curvature variations in
the hole-scale region do not affect the leading-order flow.

\section{Extensional thin-film equations for an Oldroyd-B fluid}\label{hole_setup}

In this section, we derive the hole-scale thin-film equations for an Oldroyd-B fluid as the leading-order expansion of the governing equations in $\epsilon$. To this end, we use the method of asymptotic expansions \citep{Hinch:book}, following the analysis of \citet{Howell:96} for the flow of a Newtonian fluid in a two-dimensional sheet. We also introduce several definitions and results that will be used in the analysis of the tip-scale region in $\mathsection$\ref{tip_setup}.

Based on the scalings discussed in $\mathsection$\ref{global_scaling}, we introduce the following dimensionless variables (writing $\epsilon r_0$ in place of $h_0$ to simplify the presentation),
\begin{alignat}{4}\label{dimensionless_variables}
R  & = \frac{r}{r_0},  &\qquad  R_e  & = \frac{r_e}{r_0},  &\qquad Z &  =  \frac{z}{\epsilon r_0},  &\qquad  H & = \frac{h}{\epsilon r_0},  \notag \\ 
U_R  & = \frac{u_r}{U},  &\qquad  U_Z &  =  \frac{u_z}{\epsilon U},  &\qquad  P & = \frac{p r_0}{\mu U}, &\qquad T &=  \frac{t U}{r_0}, \notag \\ 
\Sigma_{RR} &=  \frac{\sigma_{p,rr} r_0}{\mu U}, &\qquad \Sigma_{RZ} &=  \frac{\sigma_{p,rz} r_0}{\epsilon \mu U},  &\qquad  \Sigma_{ZZ} &=  \frac{\sigma_{p,zz} r_0}{\mu U},  &\qquad  \Sigma_{\theta \theta } &=  \frac{\sigma_{p,\theta \theta} r_0}{\mu U}.   
\end{alignat}
Note that we have scaled the polymeric shear stress $\sigma_{p,rz}$ to be $O(\epsilon)$ smaller than the typical stresses, consistent with the discussion in $\mathsection$\ref{global_scaling}. Also, the film shape is now denoted $\pm H(R,T)$.
It is convenient to define $\beta_s = \mu_s/\mu$ and $\beta_p = 1 - \beta_s$, which arise naturally when the problem is nondimensionalized using the total viscosity $\mu$. 

Under these scalings, the momentum equations~\eqref{momentum} in the $r$- and $z$-directions take the form
\begin{subequations}
\begin{align}
    \epsilon^2 \frac{\partial P}{\partial R} &= \beta_s \left(\frac{\partial^2 U_R}{\partial Z^2} + \epsilon^2  \frac{\partial}{\partial R}\left(\frac{1}{R}\frac{\partial}{\partial R}\left(RU_{R}\right)\right) \right) + \epsilon^2\left( \frac{\partial \Sigma_{RR}}{\partial R} +  \frac{\partial \Sigma_{RZ}}{\partial Z} + \frac{1}{R}\left(\Sigma_{RR} - \Sigma_{\theta \theta}\right)\right), \label{r_momentum} \\
    \frac{\partial P}{\partial Z} &= \beta_s \left(\frac{\partial^2 U_Z}{\partial Z^2}+\frac{\epsilon^2}{R}\frac{\partial}{\partial R}\left(R\frac{\partial U_Z}{\partial R} \right)\right)+ \epsilon^2 \left(\frac{\partial \Sigma_{RZ}}{\partial R} +  \frac{\Sigma_{RZ}}{R}\right)+\frac{\partial \Sigma_{ZZ}}{\partial Z}. \label{z_momentum}
\end{align}
\end{subequations}
Applying the stress boundary conditions~\eqref{dynamic_bc} in both the normal and tangential directions at $Z = \pm H(R,T)$, we obtain
\begin{subequations}
\begin{align}
      -P + 2\beta_s \frac{\partial U_Z}{\partial Z} + \Sigma_{ZZ} &= O(\epsilon^2),  \label{normal_stress} \\
      \beta_s \left(\frac{\partial U_R}{\partial Z} + \epsilon^2\frac{\partial U_Z}{\partial R}\right) + \epsilon^2 \Sigma_{RZ} &= \pm\epsilon^2\frac{\partial H}{\partial R}\left(2\beta_s \left( \frac{\partial U_R}{\partial R} - \frac{\partial U_Z}{\partial Z}\right) + \Sigma_{RR} - \Sigma_{ZZ}\right) + O(\epsilon^4).\label{tangential_stress}
\end{align}    
\end{subequations}
In~\eqref{normal_stress}, the $O(\epsilon^2)$ terms arise from corrections to the normal vector and the capillary contribution (see discussion in $\mathsection$\ref{global_scaling}). 

For each variable $Q$, representing any velocity component, pressure, stress component, or the film thickness, we seek an expansion in powers of
$\epsilon^2$,
\begin{equation}\label{generic_expansion}
    Q(R,Z,T;\epsilon)
    = Q^{(0)}(R,Z,T)
    + \epsilon^2 Q^{(1)}(R,Z,T)
    + \cdots .
\end{equation}

The leading-order momentum equation~\eqref{r_momentum}, together with the leading-order tangential stress balance~\eqref{tangential_stress}, imply that the leading-order radial velocity is independent of $Z$,
\begin{equation}\label{leading_order_velocity_dependence}
    U_R^{(0)} = U_R^{(0)}(R,T),
\end{equation}
as is common in free thin-film problems. This function is determined from the solvability condition at $O(\epsilon^2)$. 

Before doing so, combining the leading-order kinematic boundary condition~\eqref{kinematic_bc} with the leading-order continuity equation~\eqref{continuity} yields an alternative expression for mass conservation along the film,
\begin{equation}\label{film_mass_conservation}
\frac{\partial H^{(0)}}{\partial T} + \frac{1}{R}\frac{\partial }{\partial R}\left(RU_R^{(0)}H^{(0)}\right) = 0.
\end{equation}
Furthermore, integrating the leading-order axial momentum equation~\eqref{z_momentum} and applying the leading-order normal stress balance~\eqref{normal_stress} yields an expression for the leading-order pressure in terms of the leading-order velocities and the leading-order axial stress,
\begin{equation}\label{leading_pressure}
    P^{(0)} = 2\beta_s\frac{\partial U_Z^{(0)}}{\partial Z}+ \Sigma^{(0)}_{ZZ}=-\frac{2\beta_s}{R}\frac{\partial}{\partial R}\left(RU^{(0)}_R\right)+ \Sigma^{(0)}_{ZZ},
\end{equation}
where the second equality follows from the continuity equation~\eqref{continuity}.

We now consider the $O(\epsilon^{2})$ terms in the radial momentum equation~\eqref{r_momentum} to find
\begin{equation}\label{first_order_r_momentum}
\beta_s \frac{\partial^2 U_R^{(1)}}{\partial Z^2} = \frac{\partial P^{(0)}}{\partial R} -  \beta_s  \frac{\partial}{\partial R}\left(\frac{1}{R}\frac{\partial}{\partial R}\left(RU^{(0)}_{R}\right)\right)  - \frac{\partial \Sigma^{(0)}_{RR}}{\partial R} - \frac{\partial \Sigma^{(0)}_{RZ}}{\partial Z} - \frac{1}{R}\left(\Sigma^{(0)}_{RR} - \Sigma^{(0)}_{\theta \theta}\right).
\end{equation}
Substituting~\eqref{leading_pressure} into~\eqref{first_order_r_momentum} and rearranging then gives
\begin{equation}\label{first_order_r_momentum_simplified}
\beta_s \frac{\partial^2 U_R^{(1)}}{\partial Z^2} +  \frac{\partial \Sigma^{(0)}_{RZ}}{\partial Z}
= -3\beta_s  \frac{\partial}{\partial R}\left(\frac{1}{R}\frac{\partial}{\partial R}\left(RU^{(0)}_{R}\right)\right)  - \frac{\partial}{\partial R}\left(\Sigma_{RR}^{(0)} - \Sigma_{ZZ}^{(0)}\right)  - \frac{1}{R}\left(\Sigma^{(0)}_{RR} - \Sigma^{(0)}_{\theta \theta}\right).
\end{equation}

At this stage, it is difficult to proceed without additional knowledge of the stresses $\Sigma^{(0)}_{RR}$, $\Sigma^{(0)}_{ZZ}$, and $\Sigma^{(0)}_{\theta \theta}$. We now show that these stresses are independent of $Z$, provided they are initially independent of $Z$. To do so, we consider the Oldroyd-B constitutive equations for the stresses~\eqref{polymer_stress} at leading order in $\epsilon$,
\begin{subequations}
\label{leading_order_stress}
\begin{align}
2\beta_p\frac{\partial U^{(0)}_R}{\partial R} &= \Sigma^{(0)}_{RR} + Wi\left(\frac{\partial \Sigma^{(0)}_{RR}}{\partial T} + U^{(0)}_R\frac{\partial \Sigma^{(0)}_{RR}}{\partial R}  + U^{(0)}_Z \frac{\partial \Sigma^{(0)}_{RR}}{\partial Z} - 2\Sigma^{(0)}_{RR}\frac{\partial U^{(0)}_R}{\partial R}\right), \\
2\beta_p\frac{\partial U^{(0)}_Z}{\partial Z} &= \Sigma^{(0)}_{ZZ} + Wi\left(\frac{\partial \Sigma^{(0)}_{ZZ}}{\partial T} + U^{(0)}_R\frac{\partial \Sigma^{(0)}_{ZZ}}{\partial R}  + U^{(0)}_Z \frac{\partial \Sigma^{(0)}_{ZZ}}{\partial Z} - 2\Sigma^{(0)}_{ZZ}\frac{\partial U^{(0)}_Z}{\partial Z}\right), \\
2\beta_p\frac{U^{(0)}_R}{ R} &= \Sigma^{(0)}_{\theta \theta } + Wi\left(\frac{\partial \Sigma^{(0)}_{\theta \theta }}{\partial T} + U^{(0)}_R\frac{\partial \Sigma^{(0)}_{\theta \theta}}{\partial R}  + U^{(0)}_Z \frac{\partial \Sigma_{\theta \theta}^{(0)}}{\partial Z} - 2\Sigma^{(0)}_{\theta \theta}\frac{U^{(0)}_R}{R}\right).
\end{align}
\end{subequations}
Since $U_R^{(0)}=U_R^{(0)}(R,T)$, both $U_R^{(0)}/R$ and $\partial U_R^{(0)}/\partial R$ are independent of $Z$. Moreover, the leading-order continuity equation~\eqref{continuity} implies that $\partial U_Z^{(0)}/\partial Z$ is also independent of $Z$. With symmetry about the midplane, this gives
\begin{equation}\label{leading_order_UZ}
U_Z^{(0)}=-Z\frac{1}{R}\frac{\partial}{\partial R}\left(RU_R^{(0)}\right).
\end{equation}
Thus, the deformation rates appearing in~\eqref{leading_order_stress} are uniform across the film thickness. It follows that independence of $Z$ is preserved by the leading-order constitutive equations: if $\partial \Sigma^{(0)}_{ii}/\partial Z=0$, for $i=R,Z,\theta$, then the vertical-advection terms $U_Z^{(0)}\partial \Sigma^{(0)}_{ii}/\partial Z$ vanish, while the remaining coefficients depend only on $R$ and $T$. Thus, if the polymeric normal stresses are initially uniform across the film thickness, the leading-order dynamics preserve this structure. This is consistent with the leading-order flow being extensional rather than shear-dominated; although fluid moves in the $Z$-direction,  the deformation rates responsible for stretching the polymers are uniform, to leading order, across each cross-section of the film.

Lastly, we integrate~\eqref{first_order_r_momentum_simplified} with respect to $Z$ from $-H$ to $+H$ and apply the first-order, $O(\epsilon^2)$, tangential stress balance~\eqref{tangential_stress} to obtain an equation governing the leading-order velocity profile,
\begin{align}\label{thin_film_equation}
    0 &= 4\beta_s H^{(0)} \frac{\partial}{\partial R}\left(\frac{1}{R}\frac{\partial}{\partial R} \left(R U_R^{(0)}\right)\right) + 2\beta_s\frac{\partial H^{(0)}}{\partial R} \left( \frac{\partial U_R^{(0)}}{\partial R} +\frac{1}{R}\frac{\partial}{\partial R}\left(RU_R^{(0)}\right)\right) \notag\\
    &+ \frac{\partial}{\partial R}\left(H^{(0)}\left(\Sigma_{RR}^{(0)} - \Sigma_{ZZ}^{(0)}\right)\right)  + \frac{H^{(0)}}{R}\left(\Sigma_{RR}^{(0)} - \Sigma_{\theta \theta}^{(0)}\right).
\end{align}

Equation~\eqref{thin_film_equation}, together with the constitutive equation for the stresses~\eqref{leading_order_stress}, represents the extensional thin-film equations for an Oldroyd-B fluid with free surfaces at the boundaries. Note that these equations are distinct from the lubrication equations for a viscoelastic film on solid substrates \citep{ruangkriengsin2024translation, Oratis:25}, which are typically shear-dominated.

\section{Boundary conditions at the retracting tip}\label{tip_setup}

Thus far, we have derived the leading-order equations, in $\epsilon$, governing the flow and film thickness in the hole-scale region. However, the thin-film approximation ceases to be valid near the highly curved tip, where an inner tip-scale region must be considered. In this region, capillary stresses are dominant and drive the flow in the outer, hole-scale region.

In this section, we analyze the tip-scale region. We will not calculate its detailed flow or shape. Instead, our goal is to derive a matching condition between the tip- and hole-scale regions. This condition provides an effective boundary condition for the hole-scale problem, thereby closing its formulation.

We begin by expanding the exact hole radius $R_e(T;\epsilon)$ in powers of $\epsilon^2$,
\begin{equation}\label{hole_expansion}
R_e(T;\epsilon) = R_e^{(0)}(T) + \epsilon^2 R_e^{(1)}(T) + \cdots.
\end{equation}
The hole-scale region is thus interpreted asymptotically as the domain $R \in \left(R_e^{(0)}, \infty\right)$, with the tip located at $R = R_e^{(0)}(T)$. At leading order, the kinematic boundary condition~\eqref{kinematic_tip} at the tip is
\begin{equation}\label{leading_kinematic_tip}
\frac{dR_e^{(0)}(T)}{dT} = U_R^{(0)} \quad \text{at} \quad R = R_e^{(0)}(T) \quad \text{and} \quad Z = 0.
\end{equation}

Near the tip, the hole-scale description is no longer uniform, since the interfacial shape varies over the film-thickness scale. In the $(R,Z)$ variables, the tip region constitutes a boundary layer of thickness $O(\epsilon)$ extending away from $R = R_e^{(0)}$. To analyze this region, we introduce the stretched coordinates $(\mathcal{R}, \mathcal{Z})$,
\begin{equation}\label{stretched_variables}
\mathcal{R} = \frac{R - R_e^{(0)}}{\epsilon} \quad \text{and} \quad \mathcal{Z} = Z.
\end{equation}
Using an overline to denote variables in the tip region as functions of $(\mathcal{R}, \mathcal{Z})$, we rescale the variables from the hole-scale region, based on the discussion in $\mathsection$\ref{local_scaling}, as
\begin{alignat}{4}\label{tip_dimensionless_variables} \overline{U}_\mathcal{R} & = \frac{U_R - \frac{d R_e^{(0)}(T)}{dT}}{\epsilon}, &\qquad \overline{U}_{\mathcal{Z}} & = U_Z, &\qquad \overline{P} & = P, &\qquad \overline{H} &= H, \notag \\ \overline{\Sigma}_{\mathcal{R} \mathcal{R}} &= \Sigma_{RR}, &\qquad \overline{\Sigma}_{\mathcal{R}\mathcal{Z}} &= \epsilon \Sigma_{RZ}, &\qquad \overline{\Sigma}_{\mathcal{Z}\mathcal{Z}} &= \Sigma_{ZZ}, &\qquad \overline{\Sigma}_{\theta \theta } &= \Sigma_{\theta \theta}, \end{alignat}
where the subtraction of $\frac{d R_e^{(0)}(T)}{dT}$ in the definition of $\overline{U}_{\mathcal{R}}$ ensures that, at leading order, the velocity field is measured in the frame moving with the tip. In these inner variables, the tip region is described by the domain $\mathcal{R} \in (0,\infty)$ and $\mathcal{Z} \in (-\overline{H}, \overline{H})$. The factors of $\epsilon$ in the definitions of $\overline{U}_\mathcal{R}$ and $\overline{\Sigma}_{\mathcal{R}\mathcal{Z}}$ rescale the variables so that, in dimensional terms, the velocities scale as $\gamma/\mu$ and the stresses as $\gamma/h_0$.

In what follows, each tip-region variable $\overline{Q}$ is expanded in powers of $\epsilon$ as
\begin{equation}\label{tip_expansion}
\overline{Q}(\mathcal{R},\mathcal{Z},T ; \epsilon) = \overline{Q}^{(0)}(\mathcal{R},\mathcal{Z},T) + \epsilon \overline{Q}^{(1)}(\mathcal{R},\mathcal{Z},T)+\cdots.
\end{equation}

\subsection{Momentum balance}\label{tip_momentum}

Although we consider a three-dimensional, axisymmetric expansion of a hole, we show that the leading-order equations in the tip region reduce to an effectively two-dimensional \emph{Cartesian} problem in the $(\mathcal{R}, \mathcal{Z})$ plane. 
To justify this claim, we substitute~\eqref{stretched_variables} and~\eqref{tip_dimensionless_variables} into the $R$-direction momentum equation~\eqref{r_momentum} to obtain 
\begin{align}\label{tip_r_momentum}
    \epsilon \frac{\partial \overline{P}}{\partial \mathcal{R}} &= \beta_s \left(\epsilon \frac{\partial^2 \overline{U}_\mathcal{R}}{\partial \mathcal{Z}^2} +  \frac{\partial}{\partial \mathcal{R}}\left(\frac{1}{R_e^{(0)} + \epsilon \mathcal{R}}\frac{\partial}{\partial \mathcal{R}}\left(\left(R_e^{(0)} + \epsilon \mathcal{R}\right)\left(\frac{d R_e^{(0)}}{dT} + \epsilon \overline{U}_{\mathcal{R}}\right)\right)\right)\right) \notag \\
    &\quad + \epsilon\frac{\partial \overline{\Sigma}_{\mathcal{R}\mathcal{R}}}{\partial \mathcal{R}} + \epsilon\frac{\partial \overline{\Sigma}_{\mathcal{R}\mathcal{Z}}}{\partial \mathcal{Z}} + \frac{\epsilon^2}{R_e^{(0)} + \epsilon \mathcal{R}}\left(\overline{\Sigma}_{\mathcal{R}\mathcal{R}} - \overline{\Sigma}_{\theta \theta}\right) \notag \\
    &= \epsilon \left(\beta_s \left(\frac{\partial^2 \overline{U}_\mathcal{R}}{\partial \mathcal{Z}^2}+\frac{1}{R_e^{(0)}}\frac{\partial^2}{\partial \mathcal{R}^2}\left(R_e^{(0)}\overline{U}_{\mathcal{R}} + \mathcal{R}\frac{d R_e^{(0)}}{dT} \right)\right) + \frac{\partial \overline{\Sigma}_{\mathcal{R}\mathcal{R}}}{\partial \mathcal{R}} + \frac{\partial \overline{\Sigma}_{\mathcal{R}\mathcal{Z}}}{\partial \mathcal{Z}}\right) + O(\epsilon^2).
\end{align}
Retaining only the leading-order terms in $\epsilon$ in~\eqref{tip_r_momentum} yields
\begin{equation}\label{leading_tip_r_momentum}
    \frac{\partial \overline{P}^{(0)}}{\partial \mathcal{R}}  = \beta_s\left(\frac{\partial^2\overline{U}_\mathcal{R}^{(0)}}{\partial \mathcal{R}^2} + \frac{\partial^2\overline{U}_\mathcal{R}^{(0)}}{\partial \mathcal{Z}^2} \right) + \frac{\partial \overline{\Sigma}_{\mathcal{R}\mathcal{R}}^{(0)}}{\partial \mathcal{R}} + \frac{\partial \overline{\Sigma}_{\mathcal{R}\mathcal{Z}}^{(0)}}{\partial \mathcal{Z}}.
\end{equation}
Similarly, we apply the same procedure to the $Z$-direction momentum equation~\eqref{z_momentum} to find
\begin{equation}\label{tip_z_momentum}
     \frac{\partial \overline{P}}{\partial \mathcal{Z}} = \beta_s \left(\frac{1}{R_e^{(0)}+\epsilon \mathcal{R}}\frac{\partial}{\partial \mathcal{R}}\left(\left(R_e^{(0)}+\epsilon \mathcal{R}\right)\frac{\partial \overline{U}_\mathcal{Z}}{\partial \mathcal{R}} \right)+ \frac{\partial^2 \overline{U}_\mathcal{Z}}{\partial \mathcal{Z}^2}\right)+ \frac{\partial \overline{\Sigma}_{\mathcal{R}\mathcal{Z}}}{\partial \mathcal{R}} + \frac{\epsilon \overline{\Sigma}_{\mathcal{R}\mathcal{Z}}}{R_e^{(0)}+\epsilon \mathcal{R}}+\frac{\partial \overline{\Sigma}_{\mathcal{Z}\mathcal{Z}}}{\partial \mathcal{Z}}.
\end{equation}
At leading order in $\epsilon$, \eqref{tip_z_momentum} reduces to
\begin{equation}\label{leading_tip_z_momentum}
    \frac{\partial \overline{P}^{(0)}}{\partial \mathcal{Z}}  = \beta_s\left(\frac{\partial^2\overline{U}_\mathcal{Z}^{(0)}}{\partial \mathcal{R}^2} + \frac{\partial^2\overline{U}_\mathcal{Z}^{(0)}}{\partial \mathcal{Z}^2} \right) + \frac{\partial \overline{\Sigma}_{\mathcal{R}\mathcal{Z}}^{(0)}}{\partial \mathcal{R}} + \frac{\partial \overline{\Sigma}_{\mathcal{Z}\mathcal{Z}}^{(0)}}{\partial \mathcal{Z}}.
\end{equation}

Equations~\eqref{leading_tip_r_momentum} and~\eqref{leading_tip_z_momentum} correspond to the familiar two-dimensional (Stokes flow) momentum balance for the leading-order stress tensor,
\begin{equation}\label{reduced_momentum}
    \bar{\bnabla} \cdot \overline{\boldsymbol{\sigma}}^{(0)} = \mathbf{0},
\end{equation}
where
$\bar{\bnabla} = \left(\frac{\partial}{\partial \mathcal{R}}, \frac{\partial}{\partial \mathcal{Z}}\right)$ denotes the gradient operator in the $(\mathcal{R},\mathcal{Z})$ plane, and $\overline{\boldsymbol{\sigma}} = -\overline{P}\mathbf{I} + 2\beta_s \overline{\textbf{E}} + \overline{\boldsymbol{\Sigma}}$ is the dimensionless total stress tensor at the scale of the tip.

\subsection{Curvature}\label{tip_curvature}

We next show that, to leading order, the total curvature reduces to the meridional curvature in the $(\mathcal{R},\mathcal{Z})$ plane. For an axisymmetric surface $z = \pm h(r,t)$, the total curvature is (see figure~\ref{schematic})
\begin{equation}\label{curvature}
    \kappa =  \pm \frac{\frac{\partial^2 h}{\partial r^2 }}{\left(1 + \left(\frac{\partial h}{\partial r}\right)^2\right)^{\frac{3}{2}}} \pm \frac{\frac{\partial h}{\partial r}}{r\left(1 + \left(\frac{\partial h}{\partial r}\right)^2\right)^{\frac{1}{2}}}.
\end{equation}
We rewrite~\eqref{curvature} in terms of the tip-scale variables defined in~\eqref{stretched_variables} and~\eqref{tip_dimensionless_variables}, which yields
\begin{equation}\label{dimensionless_curvature}
\overline{\kappa} = \kappa h_0 = \pm\frac{\frac{\partial^2 \overline{H}}{\partial \mathcal{R}^2 }}{\left(1 + \left(\frac{\partial \overline{H}}{\partial \mathcal{R}}\right)^2\right)^{3/2}} \pm \frac{\epsilon\frac{\partial \overline{H}}{\partial \mathcal{R}}}{\left(R_e^{(0)}+ \epsilon \mathcal{R}\right)\left(1 + \left(\frac{\partial \overline{H}}{\partial \mathcal{R}}\right)^2\right)^{1/2}} = \pm \frac{\frac{\partial^2 \overline{H}}{\partial \mathcal{R}^2 }}{\left(1 + \left(\frac{\partial \overline{H}}{\partial \mathcal{R}}\right)^2\right)^{3/2}} + O(\epsilon).
\end{equation}
Equation~\eqref{dimensionless_curvature} shows that, at the tip scale, the azimuthal curvature is smaller than the meridional curvature by a factor of $O(\epsilon)$. It follows that the leading-order stress balance~\eqref{dynamic_bc} reduces to
\begin{equation}\label{leading_stress_balance_tip}
\overline{\textbf{n}} \cdot \overline{\boldsymbol{\sigma}}^{(0)} = K \,\overline{\textbf{n}} \quad \text{at} \quad \mathcal{Z} = \pm \overline{H}^{(0)}(\mathcal{R},T),
\end{equation}
where $\overline{\textbf{n}}$ denotes the outward unit normal to the interface $\mathcal{Z} = \pm \overline{H}^{(0)}(\mathcal{R},T)$, and $K$ is the corresponding curvature, given by
\begin{equation}
    K = \pm \frac{\frac{\partial^2 \overline{H}^{(0)}}{\partial \mathcal{R}^2}}{\left(1 + \left(\frac{\partial \overline{H}^{(0)}}{\partial \mathcal{R}}\right)^2\right)^{3/2}}.
\end{equation}

\subsection{Global force balance}\label{tip_force}

In $\mathsection$$\mathsection$\ref{tip_momentum} and \ref{tip_curvature}, we showed that the leading-order equations and their associated boundary conditions, at the tip, reduce to a two-dimensional problem in the $(\mathcal{R}, \mathcal{Z})$ plane. We now use this local problem to derive an effective boundary condition coupling the tip-scale region to the hole-scale region, following the approach of \citet{ahsan-rodolfo:26}.

For any $\Lambda > 0$, we define the truncated tip-scale region $\Omega_{\Lambda}$ as
\begin{equation}\label{truncated_domain}
   \Omega_{\Lambda} = \left\{\, (\mathcal{R}, \mathcal{Z}) \;\middle|\; 0 \le \mathcal{R} \leq \Lambda,\; -\overline{H}^{(0)}(\mathcal{R},T) \leq \mathcal{Z} \leq \overline{H}^{(0)}(\mathcal{R},T) \,\right\}.
\end{equation}
We integrate~\eqref{reduced_momentum} over the domain $\Omega_{\Lambda}$ and apply the divergence theorem to find
\begin{subequations}
\begin{align}
   \textbf{0} &= \int_{\Omega_{\Lambda}}\bar{\bnabla} \cdot \overline{\boldsymbol{\sigma}}^{(0)} \: d\Omega_{\Lambda} = \int_{\partial \Omega_{\Lambda}} \overline{\textbf{n}} \cdot \overline{\boldsymbol{\sigma}}^{(0)} \: dS \\ 
   &= \int_{0\leq \mathcal{R} \leq \Lambda,\; \mathcal{Z} = \pm \overline{H}^{(0)}} \overline{\textbf{n}} \cdot \overline{\boldsymbol{\sigma}}^{(0)} \: dS  +\int_{-\overline{H}^{(0)}(\Lambda,T)}^{\overline{H}^{(0)}(\Lambda,T)} \textbf{e}_{\mathcal{R}} \cdot \overline{\boldsymbol{\sigma}}^{(0)} \: d\mathcal{Z}, \label{integral_intermediate_step} \\
   &= \int_{0\leq \mathcal{R} \leq \Lambda,\; \mathcal{Z} = \pm \overline{H}^{(0)}} K\overline{\textbf{n}} \: dS  +\int_{-\overline{H}^{(0)}(\Lambda,T)}^{\overline{H}^{(0)}(\Lambda,T)} \textbf{e}_{\mathcal{R}} \cdot \overline{\boldsymbol{\sigma}}^{(0)} \: d\mathcal{Z}, \label{integral_final_step}
\end{align}    
\end{subequations}
where the last equality is obtained by substituting~\eqref{leading_stress_balance_tip} into~\eqref{integral_intermediate_step}. To further simplify~\eqref{integral_final_step}, we invoke the Frenet–Serret formulas for the curve $\pm \overline{H}^{(0)}$,
\begin{equation}
\frac{d\overline{\mathbf{t}}}{dS} = K\overline{\mathbf{n}},
\end{equation}
where $\overline{\mathbf{t}}$ denotes the associated unit tangent vector. This identity allows the first term in~\eqref{integral_final_step} to be evaluated explicitly as
\begin{equation}\label{Frenet_serret_integral}
    \int_{0\leq \mathcal{R} \leq \Lambda,\; \mathcal{Z} = \pm \overline{H}^{(0)}} K\overline{\textbf{n}} \: dS = \overline{\textbf{t}}\bigg|_{\mathcal{R} = \Lambda, \: \mathcal{Z} = \overline{H}^{(0)}(\Lambda)} +\overline{\textbf{t}}\bigg|_{\mathcal{R} = \Lambda, \: \mathcal{Z} = -\overline{H}^{(0)}(\Lambda)} = \frac{2}{\sqrt{1+\left(\partial \overline{H}^{(0)}/ \partial \mathcal{R}\right)^2}}\Bigg|_{\mathcal{R} = \Lambda}\textbf{e}_{\mathcal{R}}.
\end{equation}
Combining~\eqref{integral_final_step} with~\eqref{Frenet_serret_integral}, we obtain
\begin{equation}\label{effective_boundary_truncated}
    \int_{-\overline{H}^{(0)}(\Lambda)}^{\overline{H}^{(0)}(\Lambda)}  \overline{\sigma}_{\mathcal{R}\mathcal{R}}^{(0)} \: d \mathcal{Z} = -\frac{2}{\sqrt{1+\left(\partial \overline{H}^{(0)}/ \partial \mathcal{R}\right)^2}}\Bigg|_{\mathcal{R} = \Lambda}.
\end{equation}

We now take the limit $\Lambda \to \infty$ in~\eqref{effective_boundary_truncated} and apply the leading-order matching conditions between the hole-scale and tip-scale variables,
\begin{subequations}
    \begin{align}
\overline{H}^{(0)}(\mathcal{R} \to \infty)
    &= H^{(0)}\left(R \to R_e^{(0)}\right), \\
\frac{\partial \overline{H}^{(0)}}{\partial \mathcal{R}}
    (\mathcal{R} \to \infty)
    &= 0, \\
\overline{\sigma}_{\mathcal{R}\mathcal{R}}^{(0)}
    \left(\mathcal{R} \to \infty\right)
    &= \sigma^{(0)}_{RR}\left(R \to R_e^{(0)}\right).
\end{align}
\end{subequations}
These conditions yield the boundary condition at the edge of the hole,
\begin{equation}\label{effective_boundary_condition_wp}
H^{(0)}\sigma_{RR}^{(0)} = H^{(0)} \left(-P^{(0)} + 2\beta_s\frac{\partial U_R^{(0)}}{\partial R} + \Sigma_{RR}^{(0)}\right)= -1
\quad \text{as} \quad R \to R_e^{(0)}.
\end{equation}
Here, we have used the fact that $\sigma_{RR}^{(0)}$ does not depend on $Z$ (see $\mathsection$\ref{hole_setup}). We then substitute ~\eqref{leading_pressure} into~\eqref{effective_boundary_condition_wp} to conclude that
\begin{equation}\label{effective_boundary_condition}
    H^{(0)}\left(\frac{2\beta_s}{R}\frac{\partial}{\partial R}\left(RU^{(0)}_R\right)+ 2\beta_s\frac{\partial {U_R^{(0)}}}{\partial R} +\Sigma_{RR}^{(0)}- \Sigma^{(0)}_{ZZ} \right) = -1 \quad \text{as} \quad R \to R_e^{(0)}.
\end{equation}

Equation~\eqref{effective_boundary_condition} generalizes the two-dimensional analysis of~\citet{ahsan-rodolfo:26} to an axisymmetric setting, while also including non-Newtonian contributions to the stress.

\section{Hole expansion in a viscoelastic liquid sheet: thin-film model}\label{derivation_section}

In what follows, we restrict attention to the leading-order solution in $\epsilon$ and, for notational simplicity, omit the superscript $(0)$. To summarize our derivations thus far, we present below the reduced model for the velocity $U_R(R,T)$, the polymeric stress tensor $\boldsymbol{\Sigma}(R,T)$, and film thickness $H(R,T)$ in the hole-scale region.

\textit{Mass conservation}
\begin{equation}\label{derived_mass_conservation}
\frac{\partial H}{\partial T} + \frac{1}{R}\frac{\partial}{\partial R}\left(R H U_R \right) = 0.
\end{equation}

\textit{Momentum balance}
\begin{align}\label{derived_momentum_balance}
4\beta_s H \frac{\partial}{\partial R}\left(\frac{1}{R}\frac{\partial}{\partial R}(R U_R)\right) 
+ 2\beta_s \frac{\partial H}{\partial R} 
\left( \frac{\partial U_R}{\partial R} + \frac{1}{R}\frac{\partial}{\partial R}(R U_R) \right) \notag \\
\quad + \frac{\partial}{\partial R}\left(H(\Sigma_{RR} - \Sigma_{ZZ})\right) 
+ \frac{H}{R}(\Sigma_{RR} - \Sigma_{\theta\theta})&=0.
\end{align}

\textit{Oldroyd-B constitutive equations}
\begin{subequations}\label{derived_Oldroyd_B}
\begin{align}
2\beta_p \frac{\partial U_R}{\partial R} 
&= \Sigma_{RR} + Wi\left( \frac{\partial \Sigma_{RR}}{\partial T} 
+ U_R \frac{\partial \Sigma_{RR}}{\partial R} 
- 2\Sigma_{RR} \frac{\partial U_R}{\partial R} \right), \\
-2\beta_p \frac{1}{R}\frac{\partial}{\partial R}(R U_R) 
&= \Sigma_{ZZ} + Wi\left( \frac{\partial \Sigma_{ZZ}}{\partial T} 
+ U_R \frac{\partial \Sigma_{ZZ}}{\partial R} 
+ \frac{2\Sigma_{ZZ}}{R}\frac{\partial}{\partial R}(R U_R) \right), \\
2\beta_p \frac{U_R}{R} 
&= \Sigma_{\theta\theta} + Wi\left( \frac{\partial \Sigma_{\theta\theta}}{\partial T} 
+ U_R \frac{\partial \Sigma_{\theta\theta}}{\partial R} 
- \frac{2\Sigma_{\theta\theta} U_R}{R} \right).
\end{align}
\end{subequations}

\textit{Stress boundary condition at the edge}
\begin{equation}\label{derived_dynamics_BC}
   H\left( \frac{2\beta_s}{R}\frac{\partial}{\partial R}(R U_R) 
+ 2\beta_s \frac{\partial U_R}{\partial R} 
+ \Sigma_{RR} - \Sigma_{ZZ} \right) = -1  \quad \text{at} \quad  R = R_e(T)
\end{equation}

\textit{Kinematic boundary condition at the edge}
\begin{equation}\label{derived_kinematic_BC}
    \frac{dR_e}{dT} = U_R \quad \text{at} \quad  R = R_e(T)
\end{equation}

\textit{Initial and far-field conditions.}
\begin{equation}\label{derived_initial_condition}
    H(R,0)=1,\qquad
    \Sigma_{RR}(R,0)=\Sigma_{ZZ}(R,0)=\Sigma_{\theta\theta}(R,0)=0,
    \qquad
    R_e(0)=1,
\end{equation}
\begin{equation}\label{derived_far_field_condition}
    U_R\to0 \quad \text{as } \quad R\to\infty.
\end{equation}

When viscoelastic effects are absent, $Wi=0$, we recover the problem derived by \citet{Munro:Thesis} for an expanding hole in a Newtonian liquid sheet.




\subsection{Change of variables}\label{change_of_variables}

The equations above constitute a moving-boundary problem on the domain $R\in (R_e(T),\infty)$. It will be convenient for the subsequent analysis to
work instead on a stationary domain. To this end, we introduce the stretched radial coordinate
\begin{equation}\label{xi}
\xi = \frac{R}{R_e(T)}, \qquad \xi \in [1,\infty),
\end{equation}
which maps the moving edge of the hole to the fixed location $\xi=1$. For any function $F(R,T)=\mathcal{F}(\xi,T)$, the change of variables gives
\begin{subequations}\label{change_of_variables_equations} \begin{align} \frac{\partial F}{\partial T} &= \frac{\partial \mathcal{F}}{\partial T} + \frac{\partial \mathcal{F}}{\partial \xi}\frac{\partial \xi}{\partial T} = \frac{\partial \mathcal{F}}{\partial T} - \frac{\partial \mathcal{F}}{\partial \xi}\:\frac{R}{R_e^2}\:\frac{d R_e}{d T} \notag \label{time_derivative}\\ &= \frac{\partial \mathcal{F}}{\partial T} - \xi\frac{\partial \mathcal{F}}{\partial \xi}\frac{d \log R_{e}}{dT}, \\ \frac{\partial F} {\partial R} &= \frac{1}{R_e} \frac{\partial \mathcal{F}} {\partial \xi}. \label{spatial_derivative} \end{align} \end{subequations}

Note that the time derivative in \eqref{time_derivative} involves $d\log R_e/dT$, whereas
the kinematic boundary condition \eqref{derived_kinematic_BC} is written in
terms of $dR_e/dT$. We therefore introduce the rescaled radial velocity
\begin{equation}\label{new_velocity}
\tilde{U}(\xi,T) = \frac{U_R(R,T)}{R_e(T)}.
\end{equation}
In terms of $\tilde{U}$, the kinematic boundary condition becomes
\begin{equation}\label{new_kinematic}
\frac{d\log R_e}{dT} = \tilde{U}(1,T).
\end{equation}
Substituting this relation into \eqref{time_derivative} yields
\begin{equation}\label{new_time_derivative}
\left.\frac{\partial F}{\partial T}\right|_{R}
=
\frac{\partial \mathcal{F}}{\partial T}
-
\xi \tilde{U}(1,T)
\frac{\partial \mathcal{F}}{\partial \xi}.
\end{equation}

We now introduce the change of variables to the $(\xi, T)$ coordinate system, defining the film thickness and polymeric stresses as $H(R,T) = \tilde{H}(\xi,T)$ and $\boldsymbol{\Sigma}(R,T) = \tilde{\boldsymbol{\Sigma}}(\xi,T)$, respectively. Applying~\eqref{change_of_variables_equations}--\eqref{new_time_derivative}, we obtain the governing equations in a frame moving with the expanding hole:

\textit{Mass conservation}
\begin{equation}\label{new_mass_conservation}
    \frac{\partial \tilde{H}}{\partial T} - \xi\frac{\partial \tilde{H}}{\partial \xi}\tilde{U}(1,T)+ \frac{1}{\xi}\frac{\partial }{\partial \xi}\left(\xi \tilde{U}\tilde{H}\right) = 0.
\end{equation}

\textit{Momentum balance}
\begin{align}\label{new_momentum_balance}
    &4\beta_s \tilde{H} \frac{\partial}{\partial \xi}\left(\frac{1}{\xi}\frac{\partial}{\partial \xi} \left(\xi \tilde{U}\right)\right) + 2\beta_s\frac{\partial \tilde{H}}{\partial \xi} \left( \frac{\partial \tilde{U}}{\partial \xi} +\frac{1}{\xi}\frac{\partial}{\partial \xi}\left(\xi \tilde{U}\right)\right) \notag\\
    &+ \frac{\partial}{\partial \xi}\left(\tilde{H}\left(\tilde{\Sigma}_{RR} - \tilde{\Sigma}_{ZZ}\right)\right)  + \frac{\tilde{H}}{\xi}\left(\tilde{\Sigma}_{RR} - \tilde{\Sigma}_{\theta \theta}\right)=0.
\end{align}

\textit{Oldroyd-B constitutive equations}
\begin{subequations}\label{new_oldroydb}
\begin{align}
2\beta_p\frac{\partial \tilde{U}}{\partial \xi} &= \tilde{\Sigma}_{RR} + Wi\left(\frac{\partial \tilde{\Sigma}_{RR}}{\partial T}- \xi\frac{\partial \tilde{\Sigma}_{RR}}{\partial \xi}\tilde{U}(1,T) + \tilde{U}\frac{\partial \tilde{\Sigma}_{RR}}{\partial \xi}   - 2\tilde{\Sigma}_{RR}\frac{\partial \tilde{U}}{\partial \xi}\right), \\
-\frac{2\beta_p}{\xi}\frac{\partial}{\partial \xi}\left(\xi \tilde{U}\right) &= \tilde{\Sigma}_{ZZ} + Wi\left(\frac{\partial \tilde{\Sigma}_{ZZ}}{\partial T} - \xi\frac{\partial \tilde{\Sigma}_{ZZ}}{\partial \xi}\tilde{U}(1,T)+ \tilde{U}\frac{\partial \tilde{\Sigma}_{ZZ}}{\partial \xi}   + \frac{2\tilde{\Sigma}_{ZZ}}{\xi}\frac{\partial}{\partial \xi}\left(\xi \tilde{U}\right)\right), \\
\frac{2\beta_p \tilde{U}}{\xi} &= \tilde{\Sigma}_{\theta \theta } + Wi\left(\frac{\partial \tilde{\Sigma}_{\theta \theta }}{\partial T} - \xi\frac{\partial \tilde{\Sigma}_{\theta \theta}}{\partial \xi}\tilde{U}(1,T)+ \tilde{U}\frac{\partial \tilde{\Sigma}_{\theta \theta}}{\partial \xi}   - \frac{2\tilde{\Sigma}_{\theta \theta} \tilde{U}}{\xi}\right).
\end{align}    
\end{subequations}

\textit{Stress boundary condition at the edge}
\begin{equation}\label{new_dynamic_bc}
    \tilde{H}\left(\frac{2\beta_s}{\xi}\frac{\partial}{\partial \xi}\left(\xi \tilde{U}\right)+ 2\beta_s\frac{\partial \tilde{U}}{\partial \xi} +\tilde{\Sigma}_{RR}- \tilde{\Sigma}_{ZZ} \right) = -1 \quad \text{at} \quad \xi = 1.
\end{equation}

\textit{Initial and far-field conditions}
\begin{equation}\label{new_init_farfield}
\tilde{H}(\xi,0) = 1, \quad \tilde{\Sigma}_{RR}(\xi,0)=\tilde{\Sigma}_{ZZ}(\xi,0)=\tilde{\Sigma}_{\theta\theta}(\xi,0)=0, \quad \tilde{U} \to 0 \quad \text{as} \quad \xi \to \infty.
\end{equation}

Although \eqref{new_mass_conservation}--\eqref{new_init_farfield} now depend explicitly on the tip velocity $\tilde{U}(1,T)$, which must be determined as part of the problem, they do not depend on the unknown radius $R_e(T)$. Thus, to determine the hole radius, we first solve for the velocity field $\tilde{U}(\xi,T)$ and then recover $R_e(T)$ by integrating \eqref{new_kinematic} in time, subject to the initial condition $R_e(0)=1$. This gives
\begin{equation}\label{exponential_radius}
R_e(T)=\exp \left(\int_{0}^{T} \tilde{U}(1,\tilde{T})\, d\tilde{T}\right).
\end{equation}

A key result is that, if the dynamics approaches a steady state in the expanding-hole frame, \eqref{exponential_radius} shows that the radius grows exponentially in time, consistent with the observations of~\citet{Debregeas:95,Debregeas:98}.

\section{Small Weissenberg numbers $Wi\ll 1$}\label{small-wi-section}

To make further analytical progress, we consider the limit $Wi \ll 1$, in which viscoelastic effects enter as a perturbation to the Newtonian base state. In the reference frame of the hole, we expand each variable $\tilde{Q}$ in a regular power series in $Wi$:
\begin{equation}\label{small-wi-expansion}
\tilde{Q}(\xi,T;Wi)
=
\tilde{Q}^{[0]}(\xi,T) +Wi\,\tilde{Q}^{[1]}(\xi,T)+
\cdots .
\end{equation}
Here, square brackets in superscripts denote the expansion in $Wi$, distinguishing it from the parentheses used for the expansion in $\epsilon$ introduced in $\mathsection$\ref{hole_setup} and$~\mathsection$\ref{tip_setup}.

Note that satisfying the initial stress conditions in \eqref{new_init_farfield} requires an initial boundary layer on the timescale $T=O(Wi)$ \citep{ruangkriengsin:25}. This early-time regime affects only the polymer conformation; the sheet geometry, flow, and hence the retraction speed remain unchanged to leading order. Accordingly, in what follows we focus on the dynamics after this initial period, for which $T=O(1)$ or larger.

\subsection{Leading-order solution in $Wi$}\label{leading-wi-section}

At leading order in $Wi$, the conservation of mass~\eqref{new_mass_conservation} becomes
\begin{equation}\label{new_mass_conservation_leading}
    \frac{\partial \tilde{H}^{[0]}}{\partial T}
    - \xi \frac{\partial \tilde{H}^{[0]}}{\partial \xi}\tilde{U}^{[0]}(1,T)
    + \frac{1}{\xi}\frac{\partial}{\partial \xi}\left(\xi \tilde{U}^{[0]}\tilde{H}^{[0]}\right)
    = 0.
\end{equation}
Using~\eqref{new_oldroydb}, the leading-order polymeric stresses can be expressed in terms of the leading-order velocity as
\begin{equation}\label{new_oldroydb_leading}
\tilde{\Sigma}_{RR}^{[0]} = 2\beta_p \frac{\partial \tilde{U}^{[0]}}{\partial \xi},
\quad
\tilde{\Sigma}_{ZZ}^{[0]} = -\frac{2\beta_p}{\xi}\frac{\partial}{\partial \xi}\left(\xi \tilde{U}^{[0]}\right),
\quad
\text{and} \quad \tilde{\Sigma}_{\theta\theta}^{[0]} = \frac{2\beta_p \tilde{U}^{[0]}}{\xi}.
\end{equation}
Substituting~\eqref{new_oldroydb_leading} into the momentum equation~\eqref{new_momentum_balance}, and using $\beta_s + \beta_p = 1$, yields
\begin{equation}\label{new_momentum_balance_leading}
    0 = 2 \tilde{H}^{[0]} \frac{\partial}{\partial \xi}\left(\frac{1}{\xi}\frac{\partial}{\partial \xi}\left(\xi \tilde{U}^{[0]}\right)\right)
    + \frac{\partial \tilde{H}^{[0]}}{\partial \xi} \left( 2\frac{\partial \tilde{U}^{[0]}}{\partial \xi} + \frac{\tilde{U}^{[0]}}{\xi}\right).
\end{equation}
Equations~\eqref{new_mass_conservation_leading} and \eqref{new_momentum_balance_leading} are solved subject to the corresponding leading-order boundary condition at the edge~\eqref{new_dynamic_bc},
\begin{equation}\label{new_dynamic_bc_leading}
    \tilde{H}^{[0]}\left(\frac{2}{\xi}\frac{\partial}{\partial \xi}\left(\xi \tilde{U}^{[0]}\right)
    + 2\frac{\partial \tilde{U}^{[0]}}{\partial \xi}\right)
    = -1
    \quad \text{at} \quad \xi = 1,
\end{equation}
with an initially uniform film thickness $\tilde{H}^{[0]}(\xi,0) = 1$ and vanishing velocity in the far field~\eqref{new_init_farfield}.

To proceed, we first analyze the velocity at $T = 0$. Since $\tilde{H}$ is initially uniform, we have $\partial \tilde{H}^{[0]}/\partial \xi = 0$, so that the second term in~\eqref{new_momentum_balance_leading} vanishes. Consequently, the solution to~\eqref{new_momentum_balance_leading} that decays in the far field is of the form $\tilde{U}^{[0]}(\xi, 0) = C_0/\xi$ for some constant $C_0$. Substituting this velocity, together with the uniform thickness, into
\eqref{new_mass_conservation_leading}, we find that
$\partial \tilde H^{[0]}/\partial T=0$ at $T=0$. This suggests that the
initially uniform thickness may persist at later times. We therefore seek a
solution of the form
\begin{equation}\label{new_thickness_velocity_leading}
    \tilde{H}^{[0]}(\xi,T) = 1 \quad \text{and} \quad \tilde{U}^{[0]}(\xi,T) = \frac{C(T)}{\xi},
\end{equation}
for some function $C(T)$. This ansatz satisfies both
\eqref{new_mass_conservation_leading} and \eqref{new_momentum_balance_leading}
identically. The boundary condition at the edge~\eqref{new_dynamic_bc_leading}
then gives $C(T)=1/2$, and hence
\begin{equation}\label{new_velocity_leading}
    \tilde{U}^{[0]}(\xi,T) = \frac{1}{2\xi}.
\end{equation}

Before proceeding further, we note that when $Wi = 0$, the leading-order problem reduces to the purely Newtonian case. Let $R_e^{(N)}(T)$ denote the hole radius in the corresponding Newtonian problem. Using the kinematic boundary condition at the edge~\eqref{new_kinematic}, we obtain
\begin{equation}
    \frac{d\log R_e^{(N)}(T) }{dT} = \tilde{U}^{[0]}(1,T) = \frac{1}{2}.
\end{equation}
With the initial condition $R_e^{(N)}(0)  = 1$, this equation integrates to
\begin{equation}\label{exp growth}
    R_e^{(N)}(T) = e^{T/2}.
\end{equation}
Thus, in a highly viscous Newtonian film, the hole radius grows exponentially in time, and \eqref{new_thickness_velocity_leading} shows that the film thickness remains spatially uniform.

These results coincide with those derived for a Newtonian fluid by \citet{Munro:18}, who also employed a systematic asymptotic decomposition of the domain into hole- and tip-scale regions. The exponential growth law \eqref{exp growth} also agrees with the earlier results obtained by \citet{Savva:09}, who instead assumed a discontinuous initial film profile. 
\subsection{First-order correction solution in $Wi$}\label{first-wi-section}
We next examine the first-order correction in $Wi$, which captures the leading viscoelastic effects on the film thickness and hole expansion. At this order, the conservation of mass is
\begin{equation}\label{new_mass_conservation_first}
    \frac{\partial \tilde{H}^{[1]}}{\partial T} - \frac{1}{2}\xi \frac{\partial \tilde{H}^{[1]}}{\partial \xi} + \frac{1}{\xi}\frac{\partial}{\partial \xi}\left(\xi \tilde{U}^{[1]} + \frac{1}{2}\tilde{H}^{[1]}\right)=0. 
\end{equation}
In writing~\eqref{new_mass_conservation_first}, we used the result $\partial H^{[0]}/\partial \xi = 0$. The final two terms in~\eqref{new_mass_conservation_first} arise from expanding the product $\left(\tilde{U}\tilde{H}\right)^{[1]} = \tilde{U}^{[1]}\tilde{H}^{[0]} + \tilde{U}^{[0]}\tilde{H}^{[1]}$, followed by simplification using~\eqref{new_thickness_velocity_leading} and~\eqref{new_velocity_leading}.

Employing the leading-order velocity~\eqref{new_velocity_leading},  we obtain the leading-order polymeric stresses~\eqref{new_oldroydb_leading} explicitly as 
\begin{equation}\label{new_oldroydb_leading_explicit}
    \tilde{\Sigma}_{RR}^{[0]} = -\frac{\beta_p}{\xi^2}, \quad \tilde{\Sigma}_{ZZ}^{[0]} = 0, \quad \text{and} \quad \tilde{\Sigma}_{\theta \theta}^{[0]} = \frac{\beta_p}{\xi^2}.
\end{equation}
These quantities allow us to express the first-order polymeric stresses in terms of the first-order velocity by using equations~\eqref{new_oldroydb}, 
\begin{subequations}\label{new_oldroydb_first}
    \begin{align}
    \tilde{\Sigma}_{RR}^{[1]} &= 2\beta_p \frac{\partial \tilde{U}^{[1]}}{\partial \xi} -\left(\frac{\partial \tilde{\Sigma}_{RR}^{[0]}}{\partial T} -\xi \frac{\partial \tilde{\Sigma}_{RR}^{[0]}}{\partial \xi}\tilde{U}^{[0]}(1,T)+ \tilde{U}^{[0]}\frac{\partial \tilde{\Sigma}_{RR}^{[0]}}{\partial \xi}   - 2\tilde{\Sigma}_{RR}^{[0]}\frac{\partial \tilde{U}^{[0]}}{\partial \xi}\right) \notag \\ 
    &=  2\beta_p \frac{\partial \tilde{U}^{[1]}}{\partial \xi} - \left(-\frac{\beta_p}{\xi^2} + \frac{\beta_p}{\xi^4} - \frac{\beta_p}{\xi^4}\right) = 2\beta_p \frac{\partial \tilde{U}^{[1]}}{\partial \xi} + \frac{\beta_p}{\xi^2},  \\
    \tilde{\Sigma}_{ZZ}^{[1]} &= -\frac{2\beta_p}{\xi}\frac{\partial}{\partial \xi}\left(\xi \tilde{U}^{[1]}\right) - \left(\frac{\partial \tilde{\Sigma}_{ZZ}^{[0]}}{\partial T} -\xi \frac{\partial \tilde{\Sigma}_{ZZ}^{[0]}}{\partial \xi}\tilde{U}^{[0]}(1,T)+ \tilde{U}^{[0]}\frac{\partial \Sigma_{ZZ}^{[0]}}{\partial \xi}   + \frac{2\tilde{\Sigma}_{ZZ}^{[0]}}{\xi}\frac{\partial}{\partial \xi}\left(\xi \tilde{U}^{[0]}\right)\right) \notag \\ 
    &= -\frac{2\beta_p}{\xi}\frac{\partial}{\partial \xi}\left(\xi \tilde{U}^{[1]}\right), \\
    \tilde{\Sigma}_{\theta \theta}^{[1]} &= \frac{2\beta_p\tilde{U}^{[1]}}{\xi} - \left(\frac{\partial \tilde{\Sigma}_{\theta \theta }^{[0]}}{\partial T} -\xi \frac{\partial \tilde{\Sigma}_{\theta \theta}^{[0]}}{\partial \xi}\tilde{U}^{[0]}(1,T)+ \tilde{U}^{[0]}\frac{\partial \tilde{\Sigma}_{\theta \theta}^{[0]}}{\partial \xi}   - \frac{2\tilde{\Sigma}_{\theta \theta}^{[0]}\tilde{U}^{[0]}}{\xi}\right) \notag \\
    &= \frac{2\beta_p\tilde{U}^{[1]}}{\xi} -\left(\frac{\beta_p}{\xi^2} - \frac{\beta_p}{\xi^4} - \frac{\beta_p}{\xi^4}\right) = \frac{2\beta_p\tilde{U}^{[1]}}{\xi} -\frac{\beta_p}{\xi^2} + \frac{2\beta_p}{\xi^4}.
\end{align}
\end{subequations}

Substituting~\eqref{new_oldroydb_first} into~\eqref{new_momentum_balance} and expanding products at first order in $Wi$ as in~\eqref{new_mass_conservation_first}, we find
\begin{align}\label{new_momentum_balance_first}
    0 &= 4\beta_s \frac{\partial}{\partial \xi}\left(\frac{1}{\xi}\frac{\partial}{\partial \xi} \left(\xi \tilde{U}^{[1]}\right)\right) + 2\beta_s\frac{\partial \tilde{H}^{[1]}}{\partial \xi} \left( \frac{\partial \tilde{U}^{[0]}}{\partial \xi} \right) + \frac{\partial}{\partial \xi}\left(\tilde{\Sigma}_{RR}^{[1]} - \tilde{\Sigma}_{ZZ}^{[1]}\right)\notag\\
    & + \frac{\partial}{\partial \xi}\left(\tilde{H}^{[1]}\tilde{\Sigma}_{RR}^{[0]} \right)+ \frac{1}{\xi}\left(\tilde{\Sigma}_{RR}^{[1]} - \tilde{\Sigma}_{\theta \theta}^{[1]}\right)   + \frac{\tilde{H}^{[1]}}{\xi}\left(\tilde{\Sigma}_{RR}^{[0]} - \tilde{\Sigma}_{\theta \theta}^{[0]}\right) \notag \\
    &= 4 \frac{\partial}{\partial \xi}\left(\frac{1}{\xi}\frac{\partial}{\partial \xi} \left(\xi \tilde{U}^{[1]}\right)\right) -\frac{1}{\xi^2} \frac{\partial \tilde{H}^{[1]}}{\partial \xi} - \frac{2\beta_p}{\xi^5}.
\end{align}

Finally, the boundary condition at the edge~\eqref{new_dynamic_bc} at first order in $Wi$, upon using~\eqref{new_oldroydb_first}, becomes
\begin{subequations}\label{new_dynamic_bc_first}
\begin{align}
    0 &= 2\tilde{H}^{[1]}\frac{\partial \tilde{U}^{[0]}}{\partial \xi} + \frac{2}{\xi}\frac{\partial}{\partial \xi}\left(\xi \tilde{U}^{[1]}\right)+ 2\frac{\partial \tilde{U}^{[1]}}{\partial \xi} + \frac{\beta_p}{\xi^2} &\quad \text{at} \quad \xi =1, \\
    &= -\tilde{H}^{[1]}+ 2\tilde{U}^{[1]} + 4\frac{\partial\tilde{U}^{[1]}}{\partial \xi} + \beta_p &\quad \text{at} \quad \xi = 1.
\end{align}    
\end{subequations}

To facilitate the subsequent analysis, we scale out $\beta_p$ by introducing
\begin{equation}
\mathcal{U}(\xi,T) = \frac{\tilde{U}^{[1]}(\xi,T)}{\beta_p} \quad \text{and} \quad \mathcal{H}(\xi,T) = \frac{\tilde{H}^{[1]}(\xi,T)}{\beta_p}.
\end{equation}
In terms of these variables, the governing equations~\eqref{new_mass_conservation_first}
and~\eqref{new_momentum_balance_first}, together with the boundary
condition~\eqref{new_dynamic_bc_first}, take the form:

\textit{Mass conservation}
\begin{equation}\label{last_mass_conservation}
\frac{\partial \mathcal{H}}{\partial T}
- \frac{1}{2}\xi \frac{\partial \mathcal{H}}{\partial \xi}
+ \frac{1}{\xi}\frac{\partial}{\partial \xi}
\left(\xi \mathcal{U} + \frac{1}{2}\mathcal{H}\right) = 0,
\end{equation}

\textit{Momentum balance}
\begin{equation}\label{last_momentum_balance}
4 \frac{\partial}{\partial \xi}
\left(\frac{1}{\xi}\frac{\partial}{\partial \xi}(\xi \mathcal{U})\right)
- \frac{1}{\xi^2} \frac{\partial \mathcal{H}}{\partial \xi}
= \frac{2}{\xi^5},
\end{equation}

\textit{Stress boundary condition at the edge}
\begin{equation}\label{last_dynamic_bc}
\mathcal{H} = 2\mathcal{U} + 4\frac{\partial \mathcal{U}}{\partial \xi} + 1
\quad \text{at} \quad \xi = 1,
\end{equation}
with the initial condition $\mathcal{H}(\xi,0) = 0$ and the far-field condition
$\mathcal{U}(\xi \to \infty, T) = 0$. Notably, this rescaling results in an initial-boundary-value problem that is free of all parameters.

\subsubsection{Algebraic reformulation}\label{algebra_reform}

Although the initial thickness is specified in \eqref{last_mass_conservation}--\eqref{last_dynamic_bc}, no explicit initial condition is imposed on the velocity profile. This feature is consistent with the inertialess limit, in which acceleration terms have been neglected in the momentum balance \eqref{derived_momentum_balance}. We now show that $\mathcal{U}$ can be expressed explicitly in terms of $\mathcal{H}$ at each time.

Multiplying both sides of equation~\eqref{last_momentum_balance} by $\xi^2$ yields
\begin{equation}\label{last_momentum_balance_first_step}
    4 \xi^2 \frac{\partial}{\partial \xi}\left(\frac{1}{\xi}\frac{\partial}{\partial \xi} \left(\xi \mathcal{U}\right)\right) -\frac{\partial \mathcal{H}}{\partial \xi} = \frac{2}{\xi^3}.
\end{equation}
Integrating this expression with respect to $\xi$, and using the identity
\begin{equation}
\xi^2 \frac{\partial}{\partial \xi}\left(\frac{1}{\xi}\frac{\partial}{\partial \xi} \left(\xi \mathcal{U}\right)\right) = \frac{\partial}{\partial \xi} \left(\xi\frac{\partial}{\partial \xi}\left(\xi \mathcal{U}\right)\right)-2\frac{\partial}{\partial \xi}\left(\xi \mathcal{U}\right),
\end{equation}
we obtain
\begin{equation}\label{last_momentum_balance_second_step}
4\xi^2 \frac{\partial \mathcal{U}}{\partial \xi} - 4\xi \mathcal{U} - \mathcal{H} + \frac{1}{\xi^2} = f(T),
\end{equation}
where $f(T)$ remains to be determined. Evaluating~\eqref{last_momentum_balance_second_step} at $\xi = 1$ and applying the boundary condition~\eqref{last_dynamic_bc} then gives
\begin{equation}\label{last_momentum_unknown}
    f(T) = -6\mathcal{U}(1,T).
\end{equation}
We substitute~\eqref{last_momentum_unknown} into~\eqref{last_momentum_balance_second_step} and rearrange to find
\begin{equation}\label{last_momentum_balance_third_step}
    4\frac{\partial}{\partial \xi} \left(\frac{\mathcal{U}}{\xi}\right)= \frac{\mathcal{H}}{\xi^3}-\frac{6 \mathcal{U}(1,T)}{\xi^3}-\frac{1}{\xi^5}.
\end{equation}
Integrating~\eqref{last_momentum_balance_third_step} once with respect to $\xi$, and imposing the far-field decay condition on the velocity, we conclude that
\begin{equation}\label{last_momentum_balance_fourth_step}
    \mathcal{U}(\xi, T) = -\frac{\xi}{4}\int_{\xi}^{\infty}\frac{\mathcal{H}(\tilde{\xi},T)}{\tilde{\xi}^3} \: d\tilde{\xi} + \frac{3\mathcal{U}(1,T)} {4\xi} + \frac{1}{16 \xi^3}.
\end{equation}
Lastly, evaluating~\eqref{last_momentum_balance_fourth_step} at $\xi = 1$ yields a closure relation for $\mathcal{U}(1,T)$ in terms of $\mathcal{H}$,
\begin{equation}\label{last_momentum_balance_fifth_step}
    \mathcal{U}(1, T) = -\int_{1}^{\infty}\frac{\mathcal{H}(\tilde{\xi},T)}{\tilde{\xi}^3} \, d\tilde{\xi}  + \frac{1}{4}.
\end{equation}
In particular, at $T = 0$ for which $\mathcal{H}(\xi,0) = 0$, we recover the initial velocity distribution,
\begin{equation}
    \mathcal{U}(\xi,0) = \frac{1}{16}\left(\frac{3}{\xi} + \frac{1}{\xi^3}\right).
\end{equation}

It is difficult to obtain analytical solutions of \eqref{last_mass_conservation}--\eqref{last_dynamic_bc} for all time. We therefore first derive the steady-state solution analytically and then examine the time-dependent problem numerically. For the latter, we advance $\mathcal{H}$ via the transport equation~\eqref{last_mass_conservation} using a classical Runge--Kutta scheme, with $\mathcal{U}$ evaluated from $\mathcal{H}$ through~\eqref{last_momentum_balance_fourth_step} and~\eqref{last_momentum_balance_fifth_step}. This procedure avoids solving the boundary-value problem \eqref{last_momentum_balance} for $\mathcal{U}$ at each time step and improves both accuracy and convergence.

\subsubsection{Steady-state solution}
We now seek the steady-state solution, which determines the long-time behavior of the flow in the expanding-hole frame. Here, steady state refers to a time-independent solution in the $(\xi,T)$ coordinates, rather than in the original $(R,T)$ variables. Since $\xi = R/R_e(T)$ (see~\eqref{xi}), the expansion of the hole is absorbed into the coordinate transformation, so that a stationary profile in $\xi$ is compatible with a growing hole in the physical coordinates.

At steady state, the film thickness $\mathcal{H}_{\text{st}}(\xi)$ and velocity $\mathcal{U}_{\text{st}}(\xi)$ satisfy the mass conservation~\eqref{last_mass_conservation} with the time derivative set to zero,
\begin{equation}\label{steady_state_mass_conservation}
 \frac{1}{\xi}\frac{d}{d \xi}\left(\xi \mathcal{U}_{\text{st}}\right)= \frac{1}{2}\left(\xi - \frac{1}{\xi}\right)\frac{d \mathcal{H}_{\text{st}}}{d \xi}.   
\end{equation}
Using
\eqref{steady_state_mass_conservation} to eliminate $d \mathcal{H}_{\mathrm{st}}/d \xi$ from
\eqref{last_momentum_balance}  yields a second-order differential
equation for $\mathcal{U}_{\mathrm{st}}$:
\begin{equation}\label{steady_state_momentum_balance}
    2\frac{d}{d \xi}\left(\frac{1}{\xi}\frac{d}{d \xi} \left(\xi \mathcal{U}_{\text{st}}\right)\right) -\frac{1}{\xi^2(\xi^2-1)}\frac{d}{d \xi}\left(\xi \mathcal{U}_{\text{st}}\right) = \frac{1}{\xi^5}. 
\end{equation}
The general solution of~\eqref{steady_state_momentum_balance} is
\begin{equation}\label{steady_state_velocity}
    \mathcal{U}_{\text{st}}(\xi) = \frac{C_1}{4\xi}\left(2\xi^2 \Xi -\tan^{-1}(\Xi)-\tanh^{-1}(\Xi)\right)+\frac{C_2}{\xi}+\frac{2\xi}{21} - \frac{\log{\xi}}{21\xi} + \frac{1}{14\xi^3} ,
\end{equation}
where $C_1$ and $C_2$ are constants, and
\begin{equation}
    \Xi = \left(1- \frac{1}{\xi^2}\right)^{1/4} \in (0,1].
\end{equation}
The far-field condition $\mathcal{U}_{\mathrm{st}} \to 0$ as $\xi \to \infty$ then requires $C_1 = -4/21$. This condition, however, does not determine $C_2$, since $C_2/\xi\to 0$ as $\xi\to\infty$. Substituting this velocity field into~\eqref{steady_state_mass_conservation}
and integrating with respect to $\xi$ gives the steady-state film thickness
\begin{equation}\label{steady_state_thickness}
    \mathcal{H}_{\text{st}}(\xi) = \frac{2}{21}\left(\pi+6\log2-4\tan^{-1}\left(\Xi\right)-4\log(1+\Xi)-2\log\left(1+\Xi^2\right)\right)-\frac{1}{7\xi^2}.
\end{equation}
In deriving~\eqref{steady_state_thickness}, we have imposed the far-field
condition $\mathcal{H}_{\mathrm{st}}\to 0$ as $\xi\to\infty$. In the next
section, we verify numerically that this condition is satisfied and that the steady state is approached at late times. In the far-field limit, the leading $O\left(\xi^{-2}\right)$ contributions cancel, yielding
\begin{equation}\label{steady_state_thickness_far_field}
    \mathcal{H}_{\text{st}}(\xi)
    =
    \frac{1}{16\xi^4}
    +
    O\left(\xi^{-6}\right)
    \quad (\xi \gg 1) .
\end{equation}
The resulting $\xi^{-4}$ far-field decay is consistent with a free-surface deformation arising from normal-stress differences in the fluid, with the underlying flow driven by localized forcing at the rim, analogous to the analysis of the rod-climbing phenomenon~\citep{ruangkriengsin:25}.

Evaluating the steady-state film thickness at the edge gives
\begin{equation}\label{steady_state_thickness_edge}
    \mathcal{H}_{\text{st}}(1)
    =
    \frac{2\pi + 12 \log 2 - 3}{21}
    \approx 0.552.
\end{equation}
The corresponding velocity at the edge then follows from the boundary condition~\eqref{last_dynamic_bc},
\begin{equation}\label{steady_state_velocity_edge}
    \mathcal{U}_{\text{st}}(1)
    =
    \frac{1-\mathcal{H}_{\text{st}}(1)}{2}
    =
    \frac{12 - 6\log 2-\pi}{21}
    \equiv \alpha
    \approx 0.224.
\end{equation}
This allows us to determine the remaining constant,
\begin{equation}
    C_2 = \frac{17 - 12\log 2-2\pi}{42} \approx 0.057. 
\end{equation}
\subsubsection{Numerical solution} 
Thus far, we have derived the normalized first-order corrections to the steady-state film thickness and velocity, $\mathcal{H}_{\text{st}}$ and $\mathcal{U}_{\text{st}}$. This solution is independent of the initial conditions; instead, it relies on the \textit{a priori} far-field condition $\mathcal{H}_{\text{st}} \to 0$ as $\xi \to \infty$. We now verify numerically that the solution of~\eqref{last_mass_conservation}--\eqref{last_dynamic_bc}, subject to zero initial conditions, approaches this steady state at long times. Because the far-field condition is imposed in the limit $\xi \to \infty$, we take the computational domain to be 50 times the size of the hole. This finite-domain approximation introduces an error of approximately one percent, but it does not affect the main conclusions of our analysis.

\begin{figure}
    \centering
    \includegraphics[width=\columnwidth]{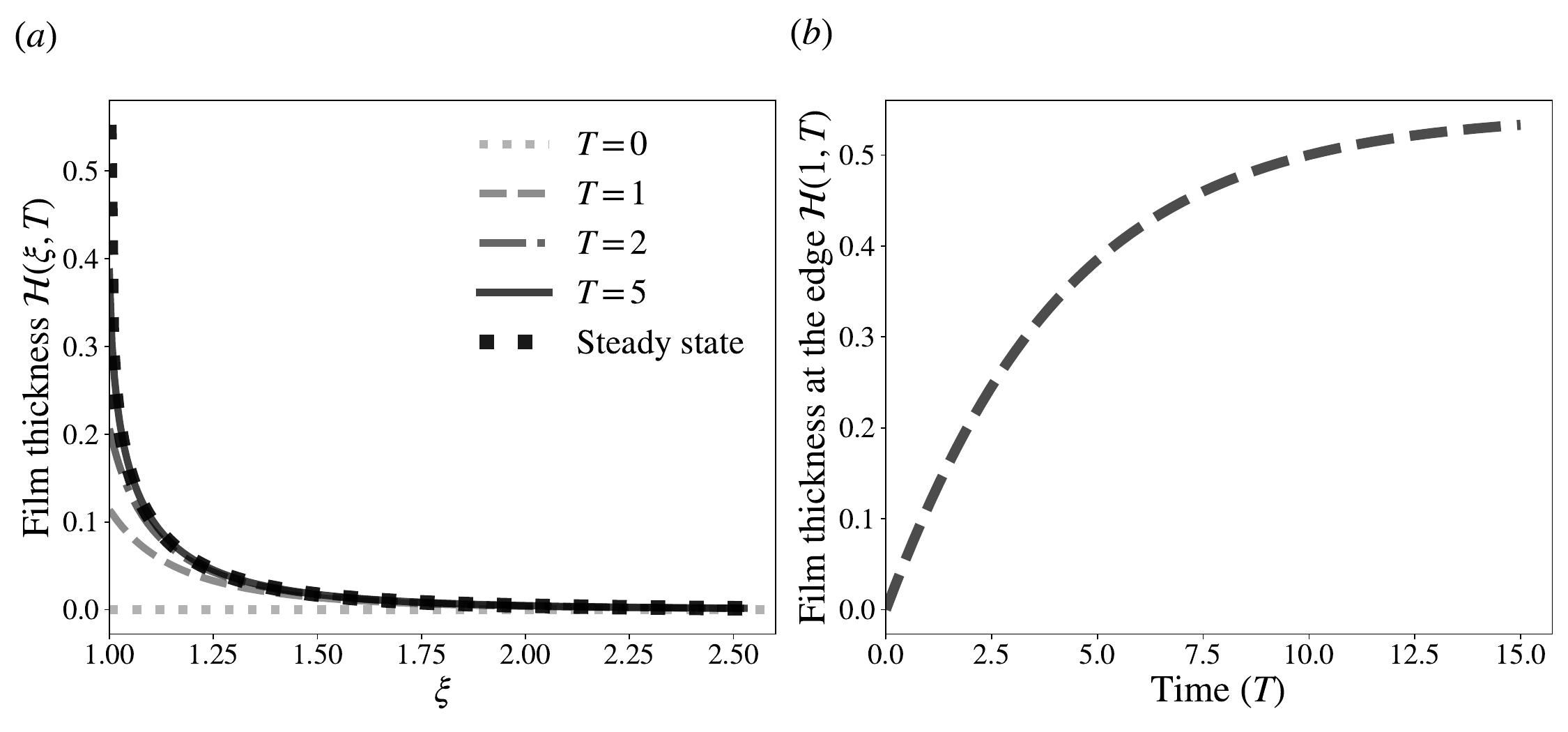}
    \caption{Time evolution of the normalized first-order correction to the film thickness,
$\mathcal{H}(\xi,T)=\tilde{H}^{[1]}(\xi,T)/\beta_p$. 
(a) Film thickness $\mathcal{H}(\xi,T)$ at $T=0,1,2$ and $5$,
compared with the analytically derived steady-state solution.
(b) Film thickness at the edge, $\mathcal{H}(1,T)$, as a function of time.}
\label{thickness_plot}
\end{figure}

\begin{figure}
    \centering
    \includegraphics[width=\columnwidth]{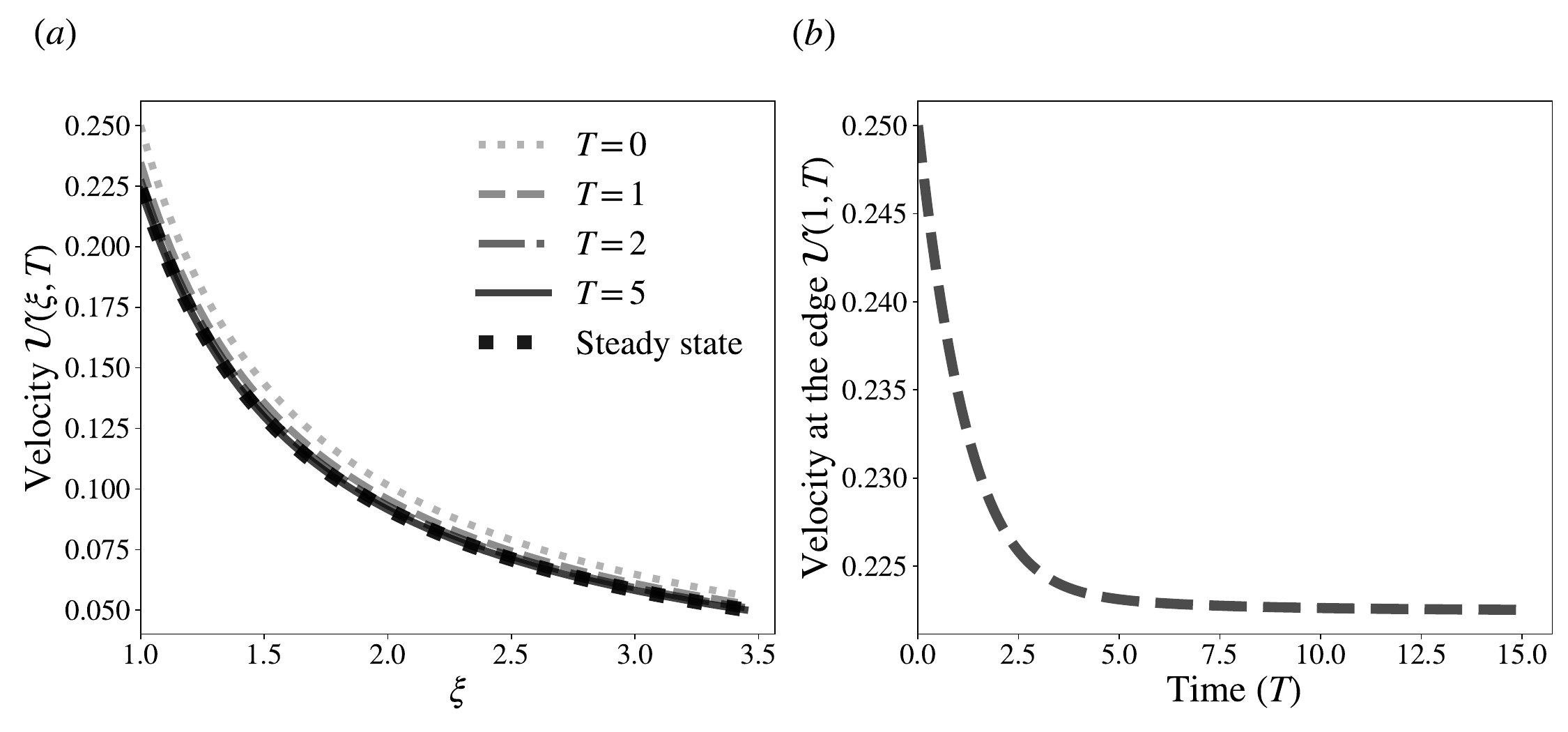}
    \caption{Time evolution of the normalized first-order correction to the velocity,
$\mathcal{U}(\xi,T)=\tilde{U}^{[1]}(\xi,T)/\beta_p$. 
(a) Velocity $\mathcal{U}(\xi,T)$ at $T=0,1,2$ and $5$,
compared with the analytically derived steady-state solution.
(b) Velocity at the edge, $\mathcal{U}(1,T)$, as a function of time.}
\label{velocity_plot}
\end{figure}

Figures~\ref{thickness_plot} and~\ref{velocity_plot} show the film thickness $\mathcal{H}$ and velocity $\mathcal{U}$ computed from the numerical implementation described in $\mathsection$\ref{algebra_reform}. The numerical solutions in both cases converge to the analytically derived steady-state solutions in~\eqref{steady_state_velocity} and~\eqref{steady_state_thickness}, thereby confirming the expected long-time behavior. Moreover, the transient film thickness reaches the steady state at $T \approx 10$, and the interface becomes nearly flat for $\xi \gtrsim 2$, indicating that the transient dynamics is localized near the retraction tip at $\xi = 1$.


Having confirmed that the solution approaches the steady state, we now determine a late-time prediction for the time evolution of the hole radius, including first-order corrections in $Wi$. Expanding the velocity $\tilde{U}(1,T)$ in \eqref{exponential_radius}, we find
\begin{equation}\label{radius_approximation_first}
R_e(T) = \exp \left(\int_{0}^{T} \tilde{U}(1,\tilde{T})\:d\tilde{T}\right)
\approx \exp \left(\frac{T}{2} + \beta_p Wi \int_{0}^{T}\mathcal{U}(1,\tilde{T}) \: d\tilde{T} \right),
\end{equation}
where the approximation retains terms through first order in $Wi$. At sufficiently late times, once $\mathcal{U}(1,T)$ has approached its steady-state value $\alpha \approx 0.224$, we have
\begin{equation}
\int_{0}^{T}\mathcal{U}(1,\tilde{T})\: d\tilde{T}
= \alpha T + \int_{0}^{T}\left(\mathcal{U}(1,\tilde{T})-\alpha\right)\: d\tilde{T}
=  \alpha T + o(T).
\end{equation}
Our numerical results suggest that $\mathcal{U}(1,T)$ approaches its steady-state value sufficiently rapidly that $\mathcal{U}(1,T)-\alpha$ is integrable, in which case the error is $O(1)$. Taking the limit of \eqref{radius_approximation_first} as $Wi\ll1$ and $T\gg1$, with $Wi T$ held fixed, we obtain
\begin{equation}\label{radius_approximation_second}
R_e(T) \approx e^{\left(\frac{1}{2}+\alpha \beta_p Wi\right)T}.
\end{equation}
This result neglects prefactors in the amplitude of $R_e(T)$ arising from the initial transient, which are asymptotically small in $Wi$. A numerical comparison between~\eqref{radius_approximation_first} and~\eqref{radius_approximation_second} confirms that the two results agree well at long times; this comparison is deferred to a later section, in figure~\ref{hole_radius}, where the final results from both the small-$Wi$ and ultra-dilute regimes are summarized and compared.
Thus, \eqref{radius_approximation_second} provides a reliable estimate for the hole radius after a short initial transient.

Our analysis demonstrates that weak viscoelasticity does not merely introduce an algebraic correction to the hole radius; rather, it increases the exponential growth rate of the hole. As the hole expands, the retracting flow stretches the polymers azimuthally, generating an azimuthal tensile stress and hence an inward elastic force that redistributes stresses through the radial force balance in the film. At the same time, the flow compresses the polymers radially, modifying the radial normal-stress difference that enters the stress balance at the tip. Together, these elastic effects lead to a stronger outward radial extensional flow. Although the accompanying film thickening partially offsets this increase, the net result is a larger edge velocity and faster hole expansion.

\section{Ultra-dilute limit $\beta_p \ll 1$}\label{ultra-dilute-section}
In $\mathsection$\ref{small-wi-section}, we developed an asymptotic solution in the limit $Wi \ll 1$, with $\beta_p$ held fixed. Although this analysis showed that viscoelasticity accelerates the expansion of the hole for sufficiently small $Wi$, this conclusion may not hold when $Wi=O(1)$. Moreover, the Oldroyd-B model is known to predict unbounded stresses in steady extensional flow above a finite critical value of $Wi$ \citep{bird:87}. In this section, we consider the alternative ultra-dilute limit $\beta_p \ll 1$, with $Wi$ fixed, to investigate the solution for larger values of $Wi$~\citep{remmelgas1999computational, moore2012weak, boyko2024flow, hinch2024fast}. This limit complements the analysis in $\mathsection$\ref{small-wi-section} and provides insight into the critical value of $Wi$ beyond which the model breaks down.

In the reference frame of the hole, we expand each variable $\tilde{Q}$ in powers of $\beta_p$:
\begin{equation}\label{small-beta-expansion}
\tilde{Q}(\xi,T;\beta_p)
=
\tilde{Q}^{\{0\}}(\xi,T) +\beta_p\,\tilde{Q}^{\{1\}}(\xi,T)+
\cdots .
\end{equation}
Curly brackets in the superscripts denote the expansion in $\beta_p$, distinguishing it from the parentheses used for the $\epsilon$ expansion introduced in $\mathsection$\ref{hole_setup} and $\mathsection$\ref{tip_setup}, and from the square brackets used for the $Wi$ expansion introduced in $\mathsection$\ref{small-wi-section}.

Under the assumption that the polymers are initially relaxed, as specified in~\eqref{new_init_farfield}, the leading-order polymeric stress satisfies the homogeneous form of~\eqref{new_oldroydb}, i.e., the equation obtained by setting the left-hand side to zero, with zero initial data and therefore vanishes identically. Consequently, the polymeric stress does not contribute to the leading-order momentum balance, and the leading-order velocity and film thickness coincide with those of the Newtonian problem analyzed in~$\mathsection$\ref{leading-wi-section}. We omit this repeated derivation and write
\begin{subequations}
\label{polymeric-beta-expansion}
\begin{align}
\tilde{H}(\xi,T) &= 1 + \beta_p \tilde{H}^{\{1\}}(\xi,T) + \cdots, \\
\tilde{U}(\xi,T) &= \frac{1}{2\xi} + \beta_p \tilde{U}^{\{1\}}(\xi,T) + \cdots, \\
\tilde{\boldsymbol{\Sigma}}(\xi,T) &= \beta_p \tilde{\boldsymbol{\Sigma}}^{\{1\}}(\xi,T) + \cdots. 
\end{align}
\end{subequations}

We emphasize that, in the ultra-dilute limit, this expansion relies strictly on the vanishing initial polymeric stress condition~\eqref{new_init_farfield}. If the initial polymeric stress is nonzero, the leading-order velocity and film thickness are coupled to the leading-order polymeric stress, resulting in a nontrivial leading-order solution. This complication does not arise in the $Wi \ll 1$ limit considered in~$\mathsection$\ref{small-wi-section}, because the initial stress condition affects the polymeric stress only over a short initial transient.

\subsection{First-order correction solution in $\beta_p$}

At first order, the mass conservation equation~\eqref{new_mass_conservation}, momentum balance~\eqref{new_momentum_balance}, and stress boundary condition at the edge~\eqref{new_dynamic_bc} follow from derivations analogous to those presented in~$\mathsection$\ref{first-wi-section}. We obtain:

\textit{Mass conservation}
\begin{equation}\label{mass_conservation_beta}
\frac{\partial \tilde{H}^{\{1\}}}{\partial T}
- \frac{1}{2}\left(\xi - \frac{1}{\xi} \right)\frac{\partial \tilde{H}^{\{1\}}}{\partial \xi}
+ \frac{1}{\xi}\frac{\partial}{\partial \xi}
\left(\xi \tilde{U}^{\{1\}} \right) = 0,
\end{equation}

\textit{Momentum balance}
\begin{equation}\label{momentum_balance_beta}
4 \frac{\partial}{\partial \xi}
\left(\frac{1}{\xi}\frac{\partial}{\partial \xi}\left(\xi \tilde{U}^{\{1\}}\right)\right)
- \frac{1}{\xi^2} \frac{\partial \tilde{H}^{\{1\}}}{\partial \xi}
= \frac{\partial}{\partial \xi}\left(\tilde{\Sigma}_{ZZ}^{\{1\}} -\tilde{\Sigma}_{RR}^{\{1\}}\right) + \frac{1}{\xi}\left(\tilde{\Sigma}_{\theta \theta }^{\{1\}} - \tilde{\Sigma}_{RR}^{\{1\}}\right),
\end{equation}

\textit{Stress boundary condition at the edge}
\begin{equation}\label{dynamic_bc_beta}
\tilde{H}^{\{1\}} = 2\tilde{U}^{\{1\}}+ 4\frac{\partial \tilde{U}^{\{1\}}}{\partial \xi} + 1 + \tilde{\Sigma}_{RR}^{\{1\}} -\tilde{\Sigma}_{ZZ}^{\{1\}}
\quad \text{at} \quad \xi = 1.
\end{equation}

To further simplify these equations and boundary conditions, we next solve for the first-order polymeric stresses. In the ultra-dilute limit considered here, these stresses are determined solely by the leading-order velocity. This differs from the $Wi \ll 1$ limit analyzed in $\mathsection$\ref{small-wi-section}, in which they also depend on the first-order velocity (see~\eqref{new_oldroydb_first}).
\subsubsection{Polymeric stresses}

Substituting the expansions for the velocity and polymeric stresses in~\eqref{polymeric-beta-expansion} into the Oldroyd-B constitutive equations~\eqref{new_oldroydb}, we obtain equations governing the first-order corrections to the polymeric stresses:

\begin{subequations}\label{new_oldroydb_beta}
\begin{align}
-\frac{1}{\xi^2} &= \tilde{\Sigma}_{RR}^{\{1\}} + Wi\left(\frac{\partial \tilde{\Sigma}_{RR}^{\{1\}}}{\partial T}- \frac{1}{2}\left(\xi-\frac{1}{\xi}\right)\frac{\partial \tilde{\Sigma}_{RR}^{\{1\}}}{\partial \xi} + \frac{\tilde{\Sigma}_{RR}^{\{1\}}}{\xi^2}\right), \\
0 &= \tilde{\Sigma}_{ZZ}^{\{1\}} + Wi\left(\frac{\partial \tilde{\Sigma}_{ZZ}^{\{1\}}}{\partial T}- \frac{1}{2}\left(\xi-\frac{1}{\xi}\right)\frac{\partial \tilde{\Sigma}_{ZZ}^{\{1\}}}{\partial \xi} \right), \\
\frac{1}{\xi^2} &= \tilde{\Sigma}_{\theta \theta}^{\{1\}} + Wi\left(\frac{\partial \tilde{\Sigma}_{\theta \theta}^{\{1\}}}{\partial T}- \frac{1}{2}\left(\xi-\frac{1}{\xi}\right)\frac{\partial \tilde{\Sigma}_{\theta \theta}^{\{1\}}}{\partial \xi} - \frac{\tilde{\Sigma}_{\theta \theta}^{\{1\}}}{\xi^2}\right).
\end{align}    
\end{subequations}

Equations~\eqref{new_oldroydb_beta}, for $Wi \neq 0$, are three uncoupled first-order linear advection equations with variable coefficients. Solving each equation subject to the initial condition~\eqref{new_init_farfield} by the method of characteristics yields
\begin{subequations}
\label{polymer-beta}
\begin{align}
    \tilde{\Sigma}_{RR}^{\{1\}}(\xi,T) &= \frac{-1+e^{-\left(\frac{1+Wi}{Wi}\right)T}}{(1+Wi)\xi^2}, \\ 
    \tilde{\Sigma}_{ZZ}^{\{1\}}(\xi,T)   &= 0, \\
    \tilde{\Sigma}_{\theta \theta}^{\{1\}}(\xi,T) &= \frac{\xi^2}{Wi} \int_{0}^{T} \frac{e^{-\left(\frac{1-Wi}{Wi}\right)\tilde{T}}}{\left(1+(\xi^2-1)e^{\tilde{T}}\right)^2} \: d\tilde{T}. \label{polymer-beta-thth}
\end{align}
\end{subequations}
Although the integral in~\eqref{polymer-beta-thth} may alternatively be expressed in terms of hypergeometric functions, its integral form allows
the polymeric stresses at the edge of the hole to be evaluated directly. Setting $\xi=1$ in the above expressions gives
\begin{equation}\label{polymer-beta-edge}
    \tilde{\Sigma}_{RR}^{\{1\}}(1,T) = \frac{-1+e^{-\left(\frac{1+Wi}{Wi}\right)T}}{1+Wi}, \quad \tilde{\Sigma}_{ZZ}^{\{1\}}(1,T)  = 0 , \quad \text{and} \quad   \tilde{\Sigma}_{\theta \theta}^{\{1\}}(1,T) =  \frac{1-e^{-\left(\frac{1-Wi}{Wi}\right)T}}{1-Wi}.
\end{equation}

Under the present nondimensionalization, \eqref{polymer-beta-edge} suggests that $Wi = 1$ is a critical value separating bounded and unbounded long-time behavior. For $Wi<1$, the azimuthal polymeric stress at the edge $\tilde{\Sigma}_{\theta \theta}^{\{1\}}(1,T)$ approaches the finite steady-state value $1/(1-Wi)$ as $T\to\infty$, whereas it grows linearly in time at $Wi=1$ and exponentially for $Wi>1$. This behavior reflects the assumption in the Oldroyd-B model that polymer molecules may extend indefinitely, so that sufficiently strong extensional flows can prevent the polymeric stress from reaching a finite steady value; this unphysical divergence is regularized in finitely extensible models such as FENE-P, which impose a finite maximum polymer extension. Note that, at any fixed position away from the edge ($\xi>1$), the integral in \eqref{polymer-beta-thth} converges as $T\to\infty$ for all $Wi$. Thus, as $Wi\to1^{-}$, the increasingly large polymeric stress becomes localized within a narrow neighborhood of the retracting edge, where the fluid experiences continuous azimuthal stretching as the hole expands. In what follows, we therefore restrict our analysis to $Wi<1$, for which the polymeric stresses remain uniformly bounded for all time.

We note that the critical value $Wi=1$ differs from the familiar Oldroyd-B threshold $Wi_{\mathrm{ext}}=1/2$ for the divergence of the polymeric stresses in steady extensional flow~\citep{bird:87} due to the different definitions of the Weissenberg number. Here, $Wi=\lambda\gamma/(\mu h_0)$ is based on the characteristic time scale of film retraction using the half-thickness $h_0$ as a length scale, whereas $Wi_{\mathrm{ext}} = \lambda E_{\theta\theta} $ is defined using the local azimuthal extension rate. From the Newtonian flow~\eqref{exp growth},
$
E_{\theta\theta}
=
u_r/r\big|_{r=r_e}
=
\gamma/(2\mu h_0),
$
so that $Wi_{\mathrm{ext}}=Wi/2$. Thus, $Wi=1$ corresponds to the classical Oldroyd-B threshold $Wi_{\mathrm{ext}}=1/2$.

\subsubsection{Steady-state solution}
As shown in $\mathsection$\ref{small-wi-section}, the long-time growth rate of the hole is governed by the steady-state velocity at its edge. In this section, we derive this velocity explicitly up to the first-order correction in $\beta_p$. Taking the limit $T \to \infty$ in~\eqref{polymer-beta}, we obtain the steady-state polymeric stresses:
\begin{equation}\label{polymer-beta-steady}
    \tilde{\Sigma}_{RR, \: \mathrm{st}}^{\{1\}}(\xi) = -\frac{1}{(1+Wi)\xi^2}, \quad \tilde{\Sigma}_{ZZ, \: \mathrm{st}}^{\{1\}}(\xi)  = 0 , \quad \text{and} \quad   \tilde{\Sigma}_{\theta \theta , \: \mathrm{st}}^{\{1\}}(\xi) = \frac{{}_2F_{1}\left(2 ,1 ; 2 + \frac{1}{Wi}; \frac{1}{\xi^2}\right)}{(1+Wi)\xi^2},
\end{equation}
where the subscript $\mathrm{st}$ denotes the steady state, and ${}_2F_1$ is the hypergeometric function. 

At steady state, the conservation of mass~\eqref{mass_conservation_beta} reduces to the same form as~\eqref{steady_state_mass_conservation},
\begin{equation}\label{steady_state_mass_conservation_beta}
 \frac{1}{\xi}\frac{d}{d \xi}\left(\xi \tilde{U}^{\{1\}}_{\rm{st}}\right)= \frac{1}{2}\left(\xi - \frac{1}{\xi}\right)\frac{d \tilde{H}^{\{1\}}_{\rm{st}}}{d \xi}.  
\end{equation}
Substituting~\eqref{polymer-beta-steady} and~\eqref{steady_state_mass_conservation_beta} into the momentum equation~\eqref{momentum_balance_beta} yields an inhomogeneous second-order ODE governing the first-order correction to the steady-state film thickness:
\begin{equation}\label{thickness-equation-beta}
2 \frac{d}{d \xi}
\left(\left(\xi - \frac{1}{\xi}\right)\frac{d \tilde{H}^{\{1\}}_{\rm{st}}}{d \xi}\right)
- \frac{1}{\xi^2} \frac{d \tilde{H}_{\rm{st}}^{\{1\}}}{d \xi}
= \frac{{}_2F_{1}\left(2 ,1 ; 2 + \frac{1}{Wi}; \frac{1}{\xi^2}\right)-1}{(1+Wi)\xi^3}.
\end{equation}

Equation~\eqref{thickness-equation-beta} can be integrated once with respect to $\xi$ by multiplying both sides by the integrating factor
$\left(\xi^{2}/(\xi^2-1)\right)^{1/4}$. This gives
\begin{equation}\label{thickness-beta-steady-derivative}
     \frac{d \tilde{H}_{\rm{st}}^{\{1\}}}{d \xi}(\xi)= \frac{\xi^{1/2}}{2(1+Wi)(\xi^2-1)^{3/4}} \mathlarger{\int}_{\xi}^{\infty}\frac{1-{}_2F_{1}\left(2 ,1 ; 2 + \frac{1}{Wi}; \frac{1}{s^2}\right)}{s^{5/2}(s^2-1)^{1/4}} \: ds.
\end{equation}
Here, the nondecaying homogeneous solution is excluded to ensure that the far-field condition $\tilde{H}_{\rm{st}}^{\{1\}} \to 0$ as $\xi \to \infty$ is satisfied. Integrating once more with respect to $\xi$, we obtain

\begin{equation}\label{thickness-beta-steady}
  \tilde{H}_{\rm{st}}^{\{1\}}(\xi)= -\frac{1}{2(1+Wi)} \mathlarger{\int}_{\xi}^{\infty}\frac{S^{1/2}}{(S^2-1)^{3/4}}\mathlarger{\int}_{S}^{\infty}\frac{1-{}_2F_{1}\left(2 ,1 ; 2 + \frac{1}{Wi}; \frac{1}{s^2}\right)}{s^{5/2}(s^2-1)^{1/4}} \: ds \: dS,
\end{equation}
where, as before, we assume that the first-order correction to the steady-state film thickness, $\tilde{H}_{\rm{st}}^{\{1\}}$, decays to zero in the far field. In principle, the first-order correction to the steady-state velocity can be obtained by substituting~\eqref{thickness-beta-steady-derivative} into~\eqref{steady_state_mass_conservation_beta} and integrating once. The resulting expression, however, also involves a double integral similar to that in~\eqref{thickness-beta-steady} and is difficult to interpret without numerical evaluation. We defer the visualization and further discussion of the thickness profile to $\mathsection$\ref{beta-numerical-solution}.
At the edge, $\xi=1$, the expression in~\eqref{thickness-beta-steady} can be reduced to a single integral by interchanging the order of integration using Fubini's theorem:
\begin{align}\label{thickness-beta-steady-edge}
  \tilde{H}_{\rm{st}}^{\{1\}}(1) &= -\frac{1}{2(1+Wi)} \mathlarger{\int}_{1}^{\infty}\mathlarger{\int}_{S}^{\infty}\frac{S^{1/2}}{(S^2-1)^{3/4}}\frac{1-{}_2F_{1}\left(2 ,1 ; 2 + \frac{1}{Wi}; \frac{1}{s^2}\right)}{s^{5/2}(s^2-1)^{1/4}} \: ds \: dS \notag \\
  &= -\frac{1}{2(1+Wi)} \mathlarger{\int}_{1}^{\infty}\frac{1-{}_2F_{1}\left(2 ,1 ; 2 + \frac{1}{Wi}; \frac{1}{s^2}\right)}{s^{5/2}(s^2-1)^{1/4}}\mathlarger{\int}_{1}^{s}\frac{S^{1/2}}{(S^2-1)^{3/4}} \: dS \: ds \notag \\
  &= -\frac{1}{1+Wi} \mathlarger{\int}_{1}^{\infty}\frac{1}{s^3}\left(1-{}_2F_{1}\left(2 ,1 ; 2 + \frac{1}{Wi}; \frac{1}{s^2}\right)\right){}_2F_{1}\left(\frac{1}{4} ,1 ; \frac{5}{4}; 1-\frac{1}{s^2}\right) \: ds \notag \\
  &= -\frac{1}{2(1+Wi)}\int_{0}^{1}\left (1-{}_2F_{1}\left(2 ,1 ; 2 + \frac{1}{Wi}; x\right)\right){}_2F_{1}\left(\frac{1}{4} ,1 ; \frac{5}{4}; 1-x\right) \: dx.
\end{align}

\begin{figure}
    \centering
    \includegraphics[width=0.59\columnwidth]{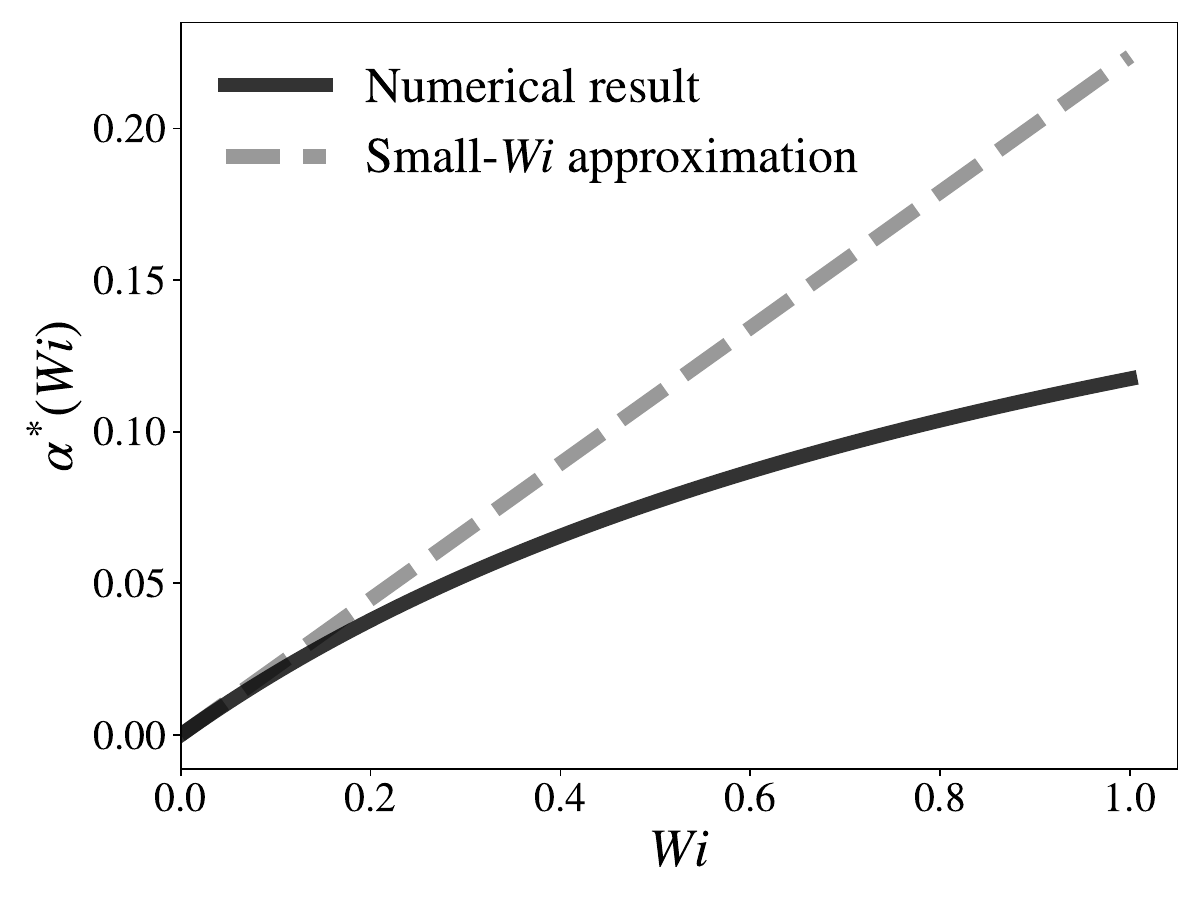}
    \caption{First-order correction to the steady-state edge velocity, $\tilde{U}^{\{1\}}_{\mathrm{st}}(1)=\alpha^{*}$, computed numerically from~\eqref{alpha_star} for $0 \leq Wi < 1$. The dashed line shows the small-$Wi$ asymptotic approximation $\alpha^{*}(Wi)\sim\alpha Wi$, where $\alpha=(12-6\log 2-\pi)/21$, as obtained in~\eqref{steady_state_velocity_edge}.}
\label{alpha_star_plot}
\end{figure}

Having determined the steady-state film thickness at the edge, we use this value to obtain the corresponding edge velocity. Evaluating~\eqref{steady_state_mass_conservation_beta} at $\xi=1$ yields $d\tilde{U}^{\{1\}}/d\xi = -\tilde{U}^{\{1\}}$ at $\xi = 1$. Combining this relation with the stress boundary condition~\eqref{dynamic_bc_beta}, we find
\begin{align}
\label{alpha_star}
    \tilde{U}^{\{1\}}_{\mathrm{st}}(1) &= \alpha^{*}(Wi) =\frac{1}{2}\left(1 - \tilde{H}^{\{1\}}_{\mathrm{st}}(1) + \tilde{\Sigma}_{RR, \: \mathrm{st}}^{\{1\}}(1) - \tilde{\Sigma}_{ZZ, \: \mathrm{st}}^{\{1\}}(1) \right) \notag \\
    &= \frac{1}{2(1+Wi)}\left(Wi +\frac{1}{2}\int_{0}^{1}\left (1-{}_2F_{1}\left(2 ,1 ; 2 + \frac{1}{Wi}; x\right)\right){}_2F_{1}\left(\frac{1}{4} ,1 ; \frac{5}{4}; 1-x\right) \: dx \right).
\end{align}
Equation~\eqref{alpha_star} is the main result of this section. As shown in the next section, this first-order correction to the steady edge velocity determines the corresponding correction to the long-time exponential growth rate of the hole, analogous to the correction in~\eqref{radius_approximation_second}. Although the expression involves a single quadrature containing hypergeometric functions, it is straightforward to evaluate numerically; the resulting plot is shown in figure~\ref{alpha_star_plot}. As $Wi\to 0$, we have $\alpha^{*}(Wi)=\alpha Wi+O(Wi^2)$, where $\alpha$ is defined in~\eqref{steady_state_velocity_edge}. In the overlapping regime $Wi\ll 1$ and $\beta_p\ll 1$, the small-$Wi$ results obtained in $\mathsection$~\ref{small-wi-section} are consistent with the small-$\beta_p$ results derived here, with the two approximations remaining in good agreement up to $Wi\approx 0.2$. As $Wi\to1^{-}$, however, we find $\alpha^{*}\approx 0.117$, which is roughly half the value predicted by the small-$Wi$ approximation $\alpha Wi$.

\subsubsection{Numerical solution}\label{beta-numerical-solution}
To verify the steady-state predictions, we numerically solve~\eqref{mass_conservation_beta}--\eqref{dynamic_bc_beta}, subject to the initial and far-field conditions~\eqref{new_init_farfield}, using the analytical polymeric stresses~$\tilde{\mathbf{\Sigma}}^{\{1\}}$ derived in~\eqref{polymer-beta}. To improve numerical convergence and avoid solving a boundary-value problem for~$\tilde{U}^{\{1\}}$, we express the velocity explicitly as a quadrature analogous to that in~\eqref{last_momentum_balance_fourth_step}, using the procedure developed in~$\mathsection$\ref{algebra_reform}: 
\begin{equation}\label{alg_reformulation_beta}
    \tilde{U}^{\{1\}}(\xi, T) = -\frac{\xi}{4}\int_{\xi}^{\infty}\frac{\mathcal{K}(\tilde{\xi},T)}{\tilde{\xi}^3} \: d\tilde{\xi} +\frac{3 \tilde{U}^{\{1\}}(1, T)}{4 \xi}.
\end{equation}
In the ultra-dilute limit considered here, however, the kernel $\mathcal{K}$ depends not only on the film thickness~$\tilde{H}^{\{1\}}$, but also on the time-dependent polymeric stresses~$\tilde{\mathbf{\Sigma}}^{\{1\}}(\xi,T)$:
\begin{equation}
    \mathcal{K}(\xi,T) = \tilde{H}^{\{1\}}(\xi,T) -1 - \xi^2\tilde{\Sigma}_{RR}^{\{1\}}(\xi,T) + \int_{1}^{\xi} \tilde{\xi}\left(\tilde{\Sigma}_{RR}^{\{1\}}(\tilde{\xi},T) + \tilde{\Sigma}_{\theta \theta}^{\{1\}}(\tilde{\xi},T)\right) \: d \tilde{\xi}.
\end{equation}
Note that the edge velocity~$\tilde{U}^{\{1\}}(1,T)$ is readily determined by evaluating~\eqref{alg_reformulation_beta} at~$\xi=1$.

The numerical problem therefore reduces to advancing the film thickness $\tilde{H}^{\{1\}}$ using the mass-conservation equation~\eqref{mass_conservation_beta} with a classical Runge--Kutta scheme, while reconstructing the velocity field~$\tilde{U}^{\{1\}}(\xi,T)$ at each time step from the quadrature~\eqref{alg_reformulation_beta}.

\begin{figure}
    \centering
    \includegraphics[width=\columnwidth]{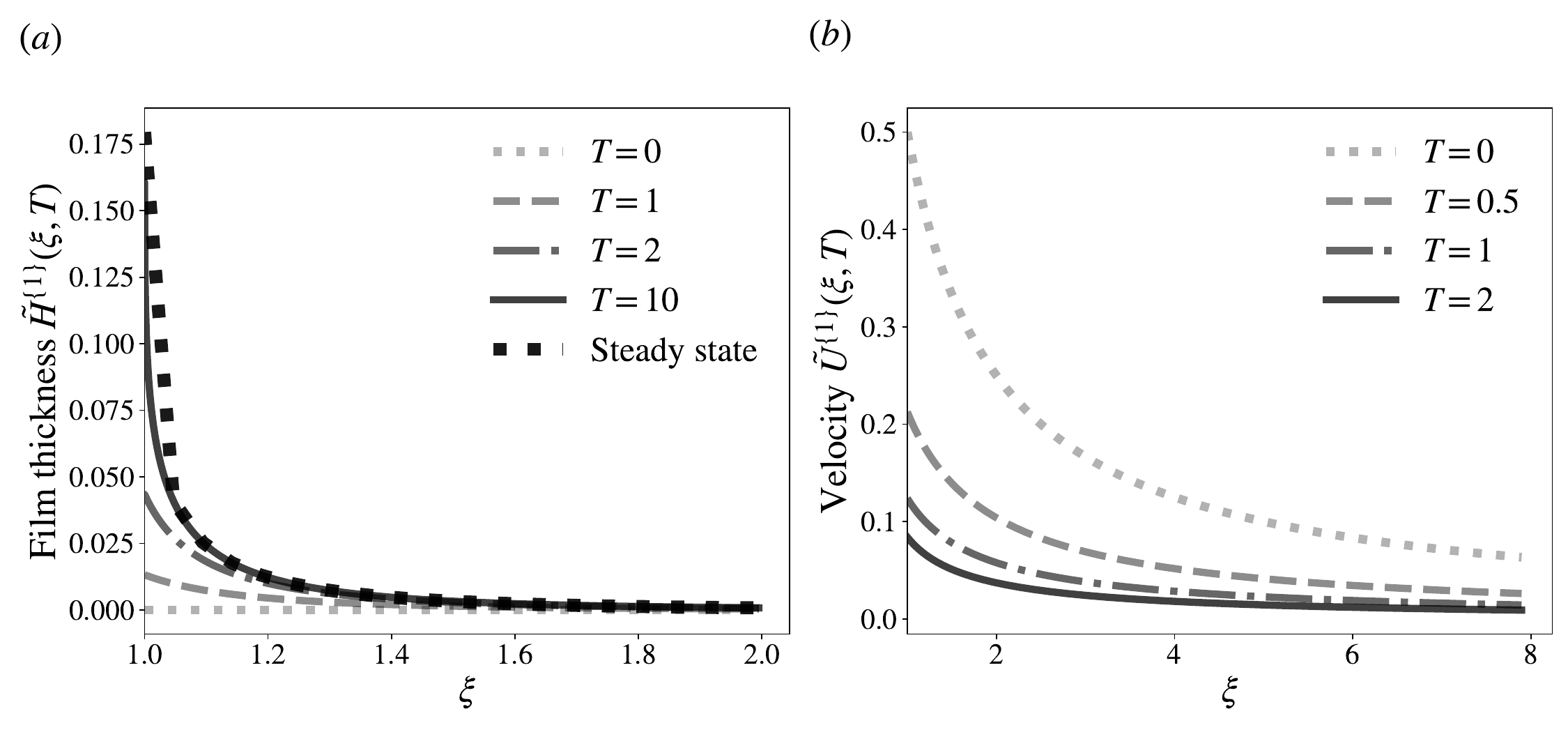}
    \caption{Time evolution of the first-order corrections to the film thickness, $\tilde{H}^{\{1\}}(\xi,T)$, and velocity, $\tilde{U}^{\{1\}}(\xi,T)$. (a) Film thickness $\tilde{H}^{\{1\}}(\xi,T)$ at $T=0,1,2,$ and $10$, compared with the analytically derived steady-state solution. (b) Velocity $\tilde{U}^{\{1\}}(\xi,T)$ at $T=0,0.5,1,$ and $2$. The velocity at the edge approaches the steady-state prediction, $\alpha^*(0.5)\approx 0.077$, as time increases. All calculations were performed using $Wi = 0.5$.}
\label{thickness_velocity_beta_plot}
\end{figure}

In figure~\ref{thickness_velocity_beta_plot}, we show the time evolution of the first-order corrections to the film thickness, $\tilde{H}^{\{1\}}(\xi,T)$, and velocity, $\tilde{U}^{\{1\}}(\xi,T)$, for the representative value $Wi=0.5$. The film-thickness correction remains localized near the retracting edge and, at long times, converges to the steady-state profile derived in~\eqref{thickness-beta-steady}, similarly to the behavior found in~$\mathsection$\ref{small-wi-section}. The velocity correction, however, exhibits a more pronounced early-time variation than in the small-$Wi$ results, before its value at the edge approaches the steady-state prediction $\alpha^*(0.5)\approx0.077$ at long times, as given by~\eqref{alpha_star}.

The apparent difference between the early-time variations of the velocity correction in the small-$Wi$ and ultra-dilute limits is not contradictory; rather, it arises from how the initial time is interpreted in the two asymptotic analyses. In the small-$Wi$ formulation, the regular expansion~\eqref{small-wi-expansion} is valid only after an initial stress-adjustment regime of duration $T=O(Wi)$~\citep{ruangkriengsin:25}. Consequently, the time $T=0$ in~$\mathsection$\ref{small-wi-section} should be interpreted as the beginning of the outer $T=O(1)$ evolution, rather than as the physical instant of rupture. By contrast, the ultra-dilute formulation retains the full $Wi$-dependence of the polymeric stresses and begins at the physical time $T=0$ with initially relaxed polymers. As such, the variation in the velocity correction in figure~\ref{thickness_velocity_beta_plot} demonstrates the two successive processes: a rapid initial adjustment of the polymeric stresses, which is omitted from the small-$Wi$ analysis, followed by the slower evolution of the film toward its steady state. 

\begin{figure}
    \centering
    \includegraphics[width=\columnwidth]{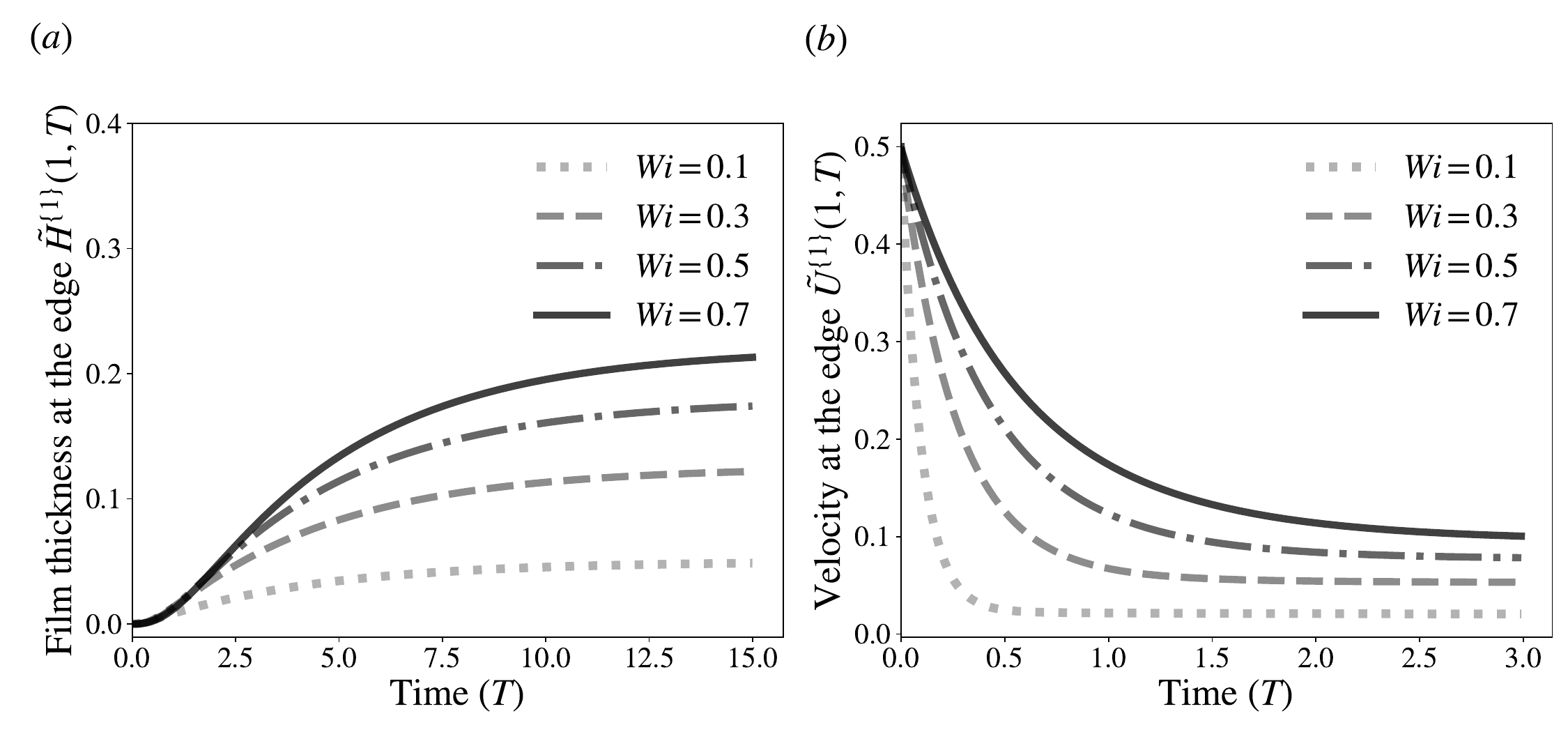}
    \caption{Time evolution of the first-order corrections to the film thickness at the edge, $\tilde{H}^{\{1\}}(1,T)$, and velocity at the edge, $\tilde{U}^{\{1\}}(1,T)$, for different values of $Wi$. (a) Film thickness at the edge $\tilde{H}^{\{1\}}(1,T)$ for $Wi = 0.1, 0.3, 0.5,$ and $0.7$. (b) Velocity at the edge $\tilde{U}^{\{1\}}(1,T)$ for $Wi=0.1,0.3,0.5,$ and $0.7$.}
\label{thickness_velocity_beta_edge_plot}
\end{figure}

This separation of timescales is further illustrated in figure~\ref{thickness_velocity_beta_edge_plot}, where both the edge thickness and velocity approach their steady values more slowly as $Wi$ increases. For smaller values of $Wi$, the initial decrease of the edge-velocity correction from $0.5$ is confined to an increasingly narrow interval $T=O(Wi)$ near $T=0$.

Having established the transient and steady behavior of the first-order
edge velocity, we now determine the corresponding evolution of the hole
radius. Following the same procedure used in $\mathsection$\ref{small-wi-section} to obtain~\eqref{radius_approximation_first}, we
substitute the ultra-dilute expansion~\eqref{small-beta-expansion} into the exact kinematic
relation~\eqref{exponential_radius} and retain terms through first order in $\beta_p$. This gives
\begin{equation}\label{radius_approximation_first_beta}
R_e(T) = \exp \left(\int_{0}^{T} \tilde{U}(1,\tilde{T})\:d\tilde{T}\right)
\approx \exp \left(\frac{T}{2} + \beta_p \int_{0}^{T}\tilde{U}^{\{1\}}(1,\tilde{T}) \: d\tilde{T} \right).
\end{equation}
Similarly, the long-time approximation follows directly from the
argument leading from~\eqref{radius_approximation_first} to~\eqref{radius_approximation_second}. For $Wi<1$, the first-order
edge-velocity correction approaches the steady value
$\alpha^*(Wi)$ obtained in~\eqref{alpha_star}, so that
\begin{equation}\label{radius_approximation_second_beta}
R_e(T) \approx e^{\left(\frac{1}{2}+\alpha^{*} \beta_p \right)T}.
\end{equation}

\begin{figure}
    \centering
    \includegraphics[width=\columnwidth]{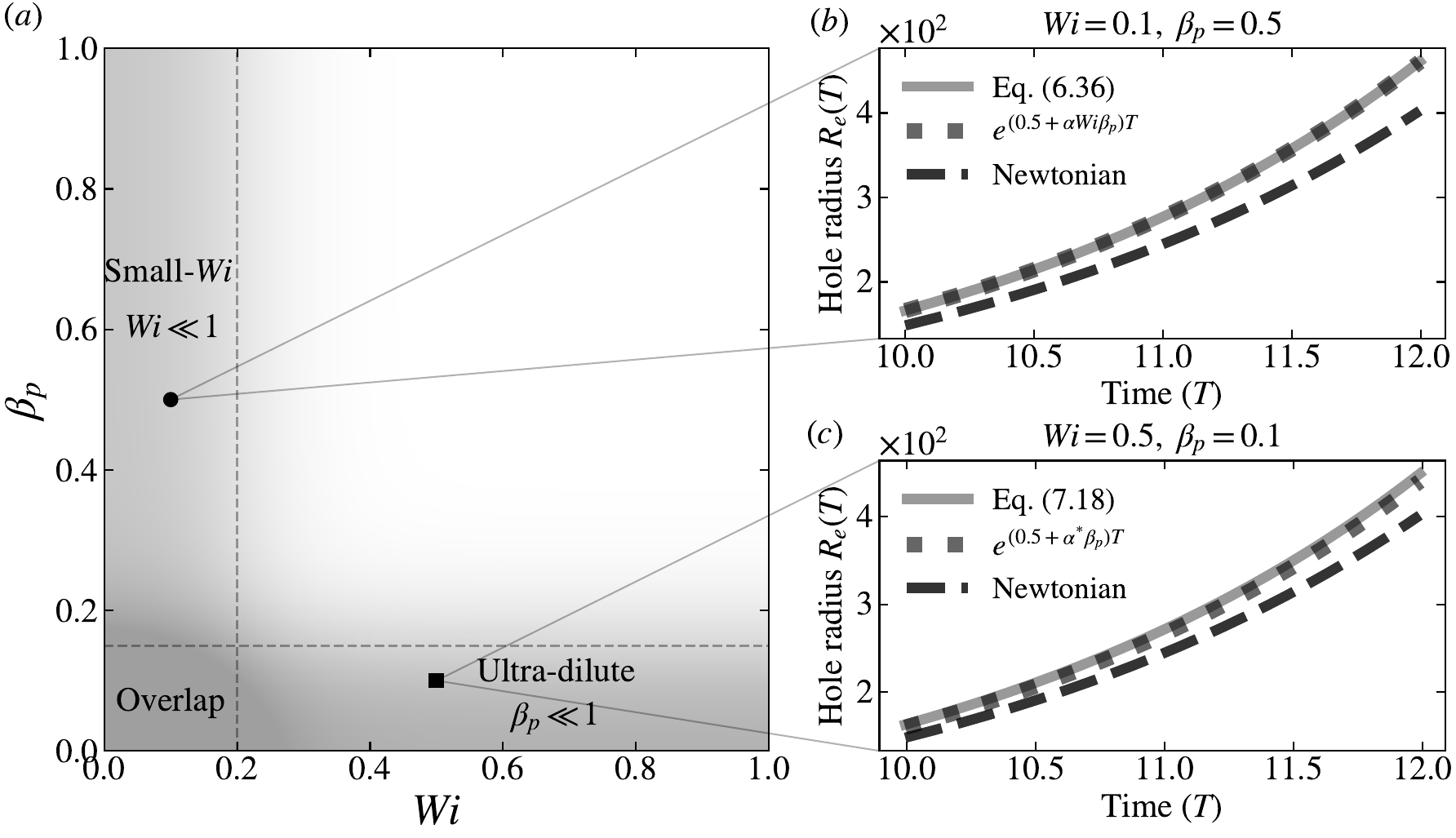}
    \caption{Summary of the two asymptotic regimes and their predictions for hole expansion. (a) Schematic parameter plane $(Wi,\beta_p)$ showing the validity regions of the small-$Wi$ expansion, $Wi \ll 1$, and the ultra-dilute expansion, $\beta_p \ll 1$, together with their overlap. Markers indicate the parameter values used in panels~(b) and~(c); the dashed boundaries are schematic rather than sharp validity thresholds. (b) Hole-radius evolution for $Wi=0.1$ and $\beta_p=0.5$, comparing the numerical transient first-order small-$Wi$ prediction~\eqref{radius_approximation_first}, the long-time approximation~\eqref{radius_approximation_second}, and the Newtonian result~\eqref{exp growth}. (c) Hole-radius evolution for $Wi=0.5$ and $\beta_p=0.1$, comparing the numerical transient first-order ultra-dilute prediction~\eqref{radius_approximation_first_beta}, the long-time approximation, and the Newtonian result~\eqref{exp growth}. The interval $10 \leq T \leq 12$ highlights the long-time agreement between the analytical and numerical predictions.}
\label{hole_radius}
\end{figure}

Thus, as in the small-$Wi$ limit considered in
$\mathsection$~\ref{small-wi-section}, the viscoelastic correction increases the exponential growth rate of the hole, thereby accelerating its retraction. Figure~\ref{hole_radius} summarizes the two complementary
asymptotic predictions developed in
$\mathsection$$\mathsection$~\ref{small-wi-section} and~\ref{ultra-dilute-section}. The
long-time analytical approximation~\eqref{radius_approximation_second_beta}
agrees well with the transient prediction
\eqref{radius_approximation_first_beta}. Moreover, since
$\alpha^*(Wi)=\alpha Wi+O(Wi^2)$ as $Wi\to0$, the ultra-dilute result
reduces to the small-$Wi$ result in their common limit. Thus, the two
distinct asymptotic approaches meet in the overlap regime $Wi \ll 1$ and $\beta_p \ll 1$ and predict
the same enhancement of the hole expansion.

\section{Conclusions}\label{conclusion}

In this work, we analyze the expansion of a hole in a highly viscous, viscoelastic liquid sheet. The problem is motivated by experiments on polymeric films~\citep{Debregeas:95, Debregeas:98, dalnoki:99}, where key features such as the absence of a localized increase in thickness near the retracting edge and the exponential growth of holes can be predicted by a purely viscous theory~\citep{Savva:09}, but the contribution of viscoelastic stresses remains unclear. We adopt the Oldroyd-B constitutive model~\citep{oldroyd:50,bird:87} to incorporate viscoelasticity into our analysis and follow the framework of~\citet{ahsan-rodolfo:26}, in which the fluid domain is decomposed into a thin-film region and a tip region coupled through an effective boundary condition.

On the scale of the hole, we derive extensional thin-film equations for an Oldroyd-B fluid based on a self-consistent asymptotic expansion that exploits the separation of length scales between the hole radius and the much smaller film thickness. In contrast to shear-dominated thin-film approximations, such as those for films on substrates~\citep{datt:22}, the leading-order flow in the present problem is extensional. Related extensional models arise in the theory of viscoelastic slender jets and filaments~\citep{forest:90,clasen:06}; however, to our knowledge, no corresponding asymptotic framework has previously been derived for a freely suspended thin viscoelastic sheet.

On the scale of the tip, we show that the leading-order problem reduces to its two-dimensional counterpart. In the present formulation, the tip region enters only through an asymptotic force balance, rather than through a solution of the full local free-boundary problem. A more complete treatment could resolve the detailed shape of the retracting edge and the local distribution of viscoelastic stresses within the tip. 

Combining the analyses on the scales of the hole and the tip, we obtain a reduced model governing the thickness and velocity of the viscoelastic film. We analyze this model in two complementary asymptotic limits. In the weakly viscoelastic limit, $Wi \ll 1$ with $\beta_p$ fixed, the hole radius grows as $e^{(0.5+\alpha Wi\beta_p )T}$, where $\alpha\approx 0.224$, at long times (see~\eqref{steady_state_velocity_edge} and~\eqref{radius_approximation_second}). In the ultra-dilute limit, $\beta_p \ll 1$ with fixed $Wi < 1$, the full $Wi$-dependence of the polymeric stresses is retained. The corresponding long-time behavior for the hole radius in this regime is $e^{(0.5+\alpha^*\beta_p )T}$, where $\alpha^*(Wi)$ is determined by numerical quadrature (see~\eqref{alpha_star} and~\eqref{radius_approximation_second_beta}). Moreover, $\alpha^{*}(Wi) = \alpha Wi + O(Wi^2)$ as $Wi \to 0$, so that the two asymptotic predictions agree in their common limit $Wi \ll 1 $ and $\beta_p \ll 1$. The ultra-dilute analysis also identifies $Wi=1$ as a critical value of the Oldroyd-B model: the azimuthal polymeric stress remains bounded at long times for $Wi<1$, but grows without bound for $Wi \geq 1$. In both regimes, the first-order correction produces a film that is thicker near the retracting edge than in the far field. These localized variations in thickness and velocity demonstrate how normal-stress differences generated by polymer stretching modify the local force balance and provide a mechanism by which viscoelasticity accelerates retraction.

Although the present work identifies a mechanism by which viscoelasticity can increase the hole growth rate, in qualitative agreement with the experiments of~\citet{Debregeas:95} and~\citet{dalnoki:99}, a quantitative comparison remains limited for several reasons. 
First, the two asymptotic analyses considered here cover the regimes $Wi \ll1 $ or $\beta_p \ll 1$, but not the general case in which both parameters are $O(1)$.
Second, we have used the Oldroyd-B constitutive model, which is more appropriate for dilute polymer solutions than for the polymer melts used in the experiments. This model provides a simple framework for isolating the effect of polymeric stresses, but neglects finite extensibility and other nonlinear rheological features, as can be seen by the predicted unbounded polymeric stresses when $Wi \geq 1$ in the ultra-dilute limit. Such effects may become important in the strongly extensional flow near the retracting edge, especially at larger values of $Wi$. It would therefore be useful to extend the present analysis to more complex constitutive laws, such as the Giesekus, FENE-P, and PTT models, which capture nonlinear rheological effects more realistically.
\backsection[Acknowledgements]{We thank Evgeniy Boyko for helpful discussions.}
\backsection[Funding]{R.B. acknowledges funding from the NSERC (Discovery Grant RGPIN-2026-06977). H.A.S. acknowledges
the support from grant no. CBET-2246791 from the United States National Science Foundation (NSF).}
\backsection[Declaration of interests]{The authors report no conflict of interest.}

\bibliographystyle{jfm}
\bibliography{refs}

\end{document}